\documentclass[twocolumn,notitlepage,aps,prb,10pt,superscriptaddress,floatfix,letterpaper,reprint
]{revtex4-2}	

\usepackage{hyperref}
\usepackage[usenames,dvipsnames]{color}
\usepackage{amsmath}
\usepackage[english]{babel}
\usepackage{amssymb}
\usepackage{graphicx}
\usepackage{epstopdf}
\usepackage[normalem]{ulem}
\usepackage{subfigure}
\usepackage{todonotes}
\usepackage{braket}
\usepackage{bbold}
\usepackage{placeins}
\usepackage[para]{threeparttable}
\usepackage{lipsum}
\usepackage{multirow}
\usepackage{bm}

\definecolor{linkcolor}{HTML}{399B03}
\definecolor{urlcolor}{HTML}{399B03}
\hypersetup{pdfstartview=FitH, linkcolor=linkcolor,urlcolor=urlcolor,colorlinks=true}
\hypersetup{
    unicode=false,          
    pdftoolbar=true,        
    pdfmenubar=true,        
    pdffitwindow=false,     
    pdfstartview={FitH},    
    pdfauthor={},         
    colorlinks=true,        
    linkcolor=blue,         
    citecolor=red,          
}

\begin{document}


\title{Testing the GFCCSD impurity solver on real materials within the self-energy embedding theory framework}

\author{Chia-Nan Yeh}%
\affiliation{%
 Department of Physics, University of Michigan, Ann Arbor, Michigan 48109, USA
}%
\author{Avijit Shee}%
\affiliation{%
 Department of Chemistry, University of Michigan, Ann Arbor, Michigan 48109, USA
}%
\author{Sergei Iskakov}
\affiliation{%
 Department of Physics, University of Michigan, Ann Arbor, Michigan 48109, USA
}%
\author{Dominika Zgid}%
\affiliation{%
 Department of Physics, University of Michigan, Ann Arbor, Michigan 48109, USA
}%
\affiliation{%
 Department of Chemistry, University of Michigan, Ann Arbor, Michigan 48109, USA
}%

\date{\today}

\begin{abstract}
We apply the Green's function coupled cluster singles and doubles (GFCCSD) impurity solver to realistic impurity problems arising for strongly correlated solids within the self-energy embedding theory (SEET) framework. We describe the details of our GFCC solver implementation, investigate its performance, and highlight potential advantages and problems on examples of impurities created during the self-consistent SEET for  antiferromagnetic MnO and paramagnetic SrMnO$_{3}$. GFCCSD provides satisfactory descriptions for weakly and moderately correlated impurities with sizes that are intractable by existing accurate impurity solvers such as exact diagonalization (ED). However, our data also shows that when correlations become strong, the singles and doubles approximation used in GFCC could lead to instabilities in searching for the particle number present in impurity problems. These instabilities appear especially severe when the impurity size gets larger and multiple degenerate orbitals with strong correlations are present. We conclude that to fully check the reliability of GFCCSD results and use them in fully {\em ab initio} calculations in the absence of experiments, a verification from a GFCC solver with higher order excitations is necessary. 
\end{abstract}

\maketitle

\section{Introduction}

At present embedding methods such as dynamical mean field theory (DMFT)~\cite{Kotliar06}, self-energy embedding theory (SEET)~\cite{Kananenka15,Zgid17,Rusakov19,Iskakov20}, or density matrix embedding theory (DMET)~\cite{Knizia12_DMET,Knizia13_DMET} for treating realistic problems reached a significant sophistication and can be routinely applied to perform calculations for systems with relatively complicated electronic structure such as transition metal oxides (e.g. NiO, MnO solids)~\cite{Ren06_NiO_LDApDMFT,Kunes07_NiO_LDApDMFT_prl,Kunes07_NiO_LDApDMFT_prb,Cui20_DMET_NiO,Zhu20_HF_CCSD,zhu2020ab,Iskakov20} or oxide perovskites (e.g. SrVO$_3$, SrMnO$_3$, and many others)~\cite{Pavarini04,SrVO3_Sekiyama04,Dang14,Chen14,Bauernfeind18,GW_EDMFT_PRM_Philipp20,Yeh20}. In these systems, in order to reach a good comparison with experiments, it is possible to embed only a small subset of orbitals present in the entire unit cell. Most frequently only the $d$-orbitals of transition metal atoms are embedded as impurity problems containing the $t_{2g}$ and $e_g$ orbitals. Such calculations are commonly performed using the LDA+DMFT formalism, where the two-body interactions describing the impurity orbitals can either be chosen to recover the experimental data or come from earlier calculations such as constrained LDA (cLDA)~\cite{Gunnarsson89}. While LDA+DMFT has an enormous computational advantage of being relatively inexpensive to calculate, in addition to a practical advantage of frequently using adjustable parameters, it suffers from some degree of double counting of electron correlation due to a lack of an explicit diagrammatic form of the DFT functional.~\cite{Dang14}.

In recent years, the rise of fully {\em ab initio} Green's function embedding methods such as GW+(E)DMFT~\cite{Biermann03} and SEET opened a possibility of performing calculations even in the absence of any experimental data without any adjustable parameters or double counting corrections.
However, if a fully {\em ab initio} calculation is required that yields a quantitative agreement with experimental data multiple challenges have to be fullfilled. First, a diagrammatic method capable of describing weak correlation has to be chosen such that it is possible to remove the double counting correction rigorously. Second, a method capable of solving an impurity problem has to be chosen. Ideally, such a method, called an impurity solver, should be systematically improvable and capable of solving a large impurity containing a large number of orbitals. These requirements are necessary to design a series of checks that allow us to confirm the validity of the results obtained even in the absence of experimental data. 
Moreover, a robust solver capable of treating a large orbital spaces certainly would deliver quantitative results and enable many new applications to compounds where both transition metal $d$-orbitals as well as $p$-orbitals of oxygen or sulphur contain a significant amount of strong correlation and should be treated within a single impurity. 

Recently, the coupled cluster (CC) method~\cite{Bartlett07} at the level of singles and doubles (CCSD) was adapted to work as a solver in conjunction with DMFT \cite{ZhuPRB19,Zhu20_HF_CCSD,zhu2020ab} and SEET \cite{Avijit19} schemes. We will call this solver a Green's function coupled cluster (GFCC) solver.
GFCC at the CCSD level (GFCCSD) is capable of treating impurities with even a couple hundred of orbitals.
While the CC method originated in the nuclear physics community, early on it was adapted by quantum chemists where it was primarily used to treat weakly correlated molecular systems. The early tests of the CC solver were performed on the impurities coming from the 1D and 2D Hubbard models~\cite{Avijit19,ZhuPRB19}, small molecular systems~\cite{Avijit19} and limited solids~\cite{zhu2020ab}. 
These tests demonstrated that the GFCCSD solver can be used in weakly to moderately strongly correlated situations. These results were somewhat surprising since CCSD, in a molecular setting, is commonly thought as being suitable only for weakly correlated problems. The success of GFCCSD for impurity problems in the 1D and 2D Hubbard case can be rationalized by noticing that CCSD is exact for a two-electron system. Consequently, for smaller impurity problems without a significant degeneracy near the Fermi level, one can expect very accurate results since it is likely that very few orbitals have significant partial occupations.  Both in cases of the 1D and 2D Hubbard model, GFCCSD was very successful where only two impurity orbitals were significantly partially occupied. Consequently, for realistic problems, it is natural to expect that the performance of the GFCCSD impurity solver may somewhat worsen.

Our motivation in this paper is to examine the results of the GFCCSD solver in detail for a set of impurities that are obtained when treating realistic problems such as MnO and SrMnO$_3$ solids. We choose these systems since the MnO solid is a prototypical antiferromagnetic (AFM) insulator, while the paramagnetic (PM) SrMnO$_3$ solid is metallic at the $GW$ level and only higher post-$GW$ correlations are opening its gap and recovering its insulating character observed experimentally~\cite{Yeh20}. We believe that such realistic compounds are more illustrative of the GFCCSD performance than the test performed on sparse model Hamiltonians.
Subsequently, we examine how the results obtained from the GFCCSD treatment of these realistic impurity problems are influencing the convergence of the self-consistency loop and spectra obtained in the SEET procedure. We focus our discussion of the performance of the GFCCSD solver on analyzing steps that are necessary for its execution and on analyzing its advantages and drawbacks. 
Moreover, any solver requires a prescription of how to use it in order to obtain a controlled and if possible an improvable set of results. We focus our discussion on establishing such a procedure and we list major requirements.

This paper is organized as follows. In Sec.~\ref{sec:Methods}, we start our discussion from briefly listing the requirements for the SEET self-consistency and an in detail description of the GFCC solver when used for the treatment of the impurity problem. Here, we particularly focus our considerations on explaining finding a proper number of particles present in the impurity problem when using the GFCC solver.
In Sec.~\ref{Sec:Results}, we compare the self-energies obtained from ED and GFCCSD solvers and discuss the final resulting spectral functions for MnO and SrMnO$_3$ solids.
Again, we pay a considerable attention to explaining the intricacies of the CC type solvers and we discuss possible difficulties resulting from their use.
Finally, in Sec.~\ref{sec:Conclusions}, we present an extensive discussion of the GFCCSD accuracy and we make recommendations concerning its controlled and systematic use for real material calculations.

\section{Methods}\label{sec:Methods}
We explore the capability of the GFCCSD impurity solver~\cite{Avijit19} (see Sec.~\ref{subsec:CCSD} for details) for \emph{ab initio} impurity problems constructed during an embedding procedure for AFM MnO and PM SrMnO$_{3}$ solids. 
The impurity problems are defined during the execution of SEET~\cite{Iskakov20,Yeh20}. Here, SEET employs the $GW$ approximation as the weakly correlated method and the GFCCSD solver as the impurity solver for strongly correlated orbitals. We shall call this variant of the SEET execution SEET($GW$/CCSD). 
In order to analyze results from the GFCCSD solver, SEET with the exact diagonalization (ED)~\cite{Caffarel94,ED_Sergei18} impurity solver, referred as SEET($GW$/ED), is performed whenever the impurity size is possible to be handled by ED. 

SEET by constructions requires no adjustable parameters, a projection to Wannier orbitals or downfolding to a low-energy effective model, and is free from the double counting correction error present in the density functional theory (DFT)~\cite{Kohn65_DFT} plus DMFT~\cite{Dang14}. 
Instead, in SEET, the \emph{ab initio} impurity Hamiltonian is constructed using bare interactions and all the non-local screening effects are treated at the $GW$ level in the self-energy embedding procedure. 
When a full self-consistency is achieved, SEET by construction yields a conserving approximation to the Luttinger-Ward functional $\Phi$~\cite{Luttinger60,Baym61,Baym62} which is thermodynamically consistent. 
However, since GFCCSD itself is not a conserving approximation~\cite{Avijit19}, SEET($GW$/CCSD) is not a conserving approximation either. 
The details of the SEET implementation used here can be found in Ref.~\cite{Iskakov20,Yeh20}. 

\begin{figure}[htp]
\includegraphics[width=0.45\textwidth]{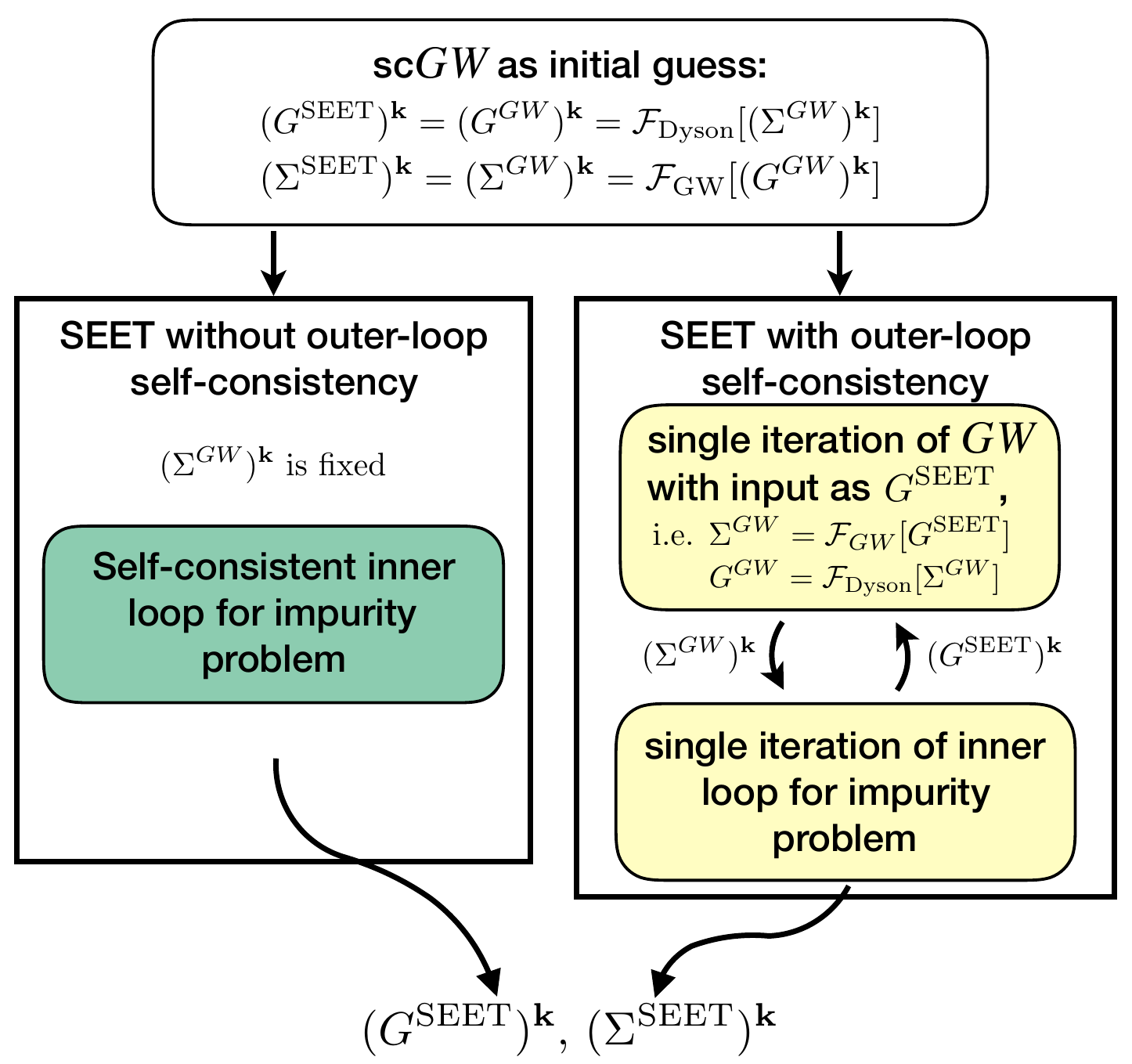}
\caption{Workflows of SEET with and without the outer-loop self-consistency. The details of the inner loop are shown in Fig.~\ref{fig:ccsd_solver}. $\mathcal{F}_{\text{Dyson}}$ represents the Dyson equation solver which is a functional of self-energy $\Sigma$ and $\mathcal{F}_{GW}$ is the $GW$ solver which is a functional of Green's function $G$. Details of $\mathcal{F}_{\text{Dyson}}$ and $\mathcal{F}_{GW}$ can be found in Ref.~\cite{Iskakov20}.}
\label{fig:SEET_workflow}
\end{figure}

For computational convenience, the outer-loop self-consistency in SEET is omitted for the AFM MnO solid~\cite{Iskakov20}. 
The main effect of the outer-loop self-consistency is to relax weakly correlated orbitals in the presence of strong correlation coming from impurity orbitals. Therefore, the omission of the outer-loop self-consistency is justified when most correlations in a solid are qualitatively captured by the weakly correlated method, e.g. self-consistent $GW$ (sc$GW$) in the present work. 
As we observed in Ref.~\cite{Iskakov20}, local self-energy corrections from SEET to correlated orbital subspaces of MnO are responsible for quantitative renormalizations between quasiparticle peaks and the satellite peaks. 
However, this is not the case for the PM SrMnO$_{3}$ solid, where sc$GW$ predicts a qualitatively incorrect metallic state~\cite{Yeh20}. Consequently, for the PM SrMnO$_{3}$ solid, we perform the outer loop self-consistency and update the weakly correlated orbital description.
The workflows for both SEET with and without the  outer-loop self-consistency are shown in Fig.~\ref{fig:SEET_workflow}. 

\subsection{Coupled Cluster Green's Function Solver}\label{subsec:CCSD}

The GFCCSD solver consists of two parts, as has been shown in Fig. \ref{fig:ccsd_solver}. The specific details of these two parts are given below.

\subsubsection{Particle sector search}

In order to define the impurity problem completely, besides the Hamiltonian matrix elements, we need to know the number of particles present for a given impurity Hamiltonian. While the knowledge of the Hamiltonian of the full problem allows us to define the impurity Hamiltonian, from the knowledge of the number of particles in the full problem we cannot immediately and \textit{a priori} define how many particles are present in the impurity problem. Therefore for the impurity problem, we scan through all possible particle numbers to determine which one yields the minimum energy while keeping the chemical potential fixed at the value found for the whole problem. 
Note that in the current work although $GW$ is conducted at a finite temperature, we consider the zero-temperature limit for the GFCCSD solver where only particle sectors with the lowest energy are used in the Green's function construction. It turns out to be a reasonable approximation since the Boltzmann factors corresponding to the first excited states are often found to be negligible in our calculations. 
\begin{figure}
    \centering
    \includegraphics[width=0.45\textwidth]{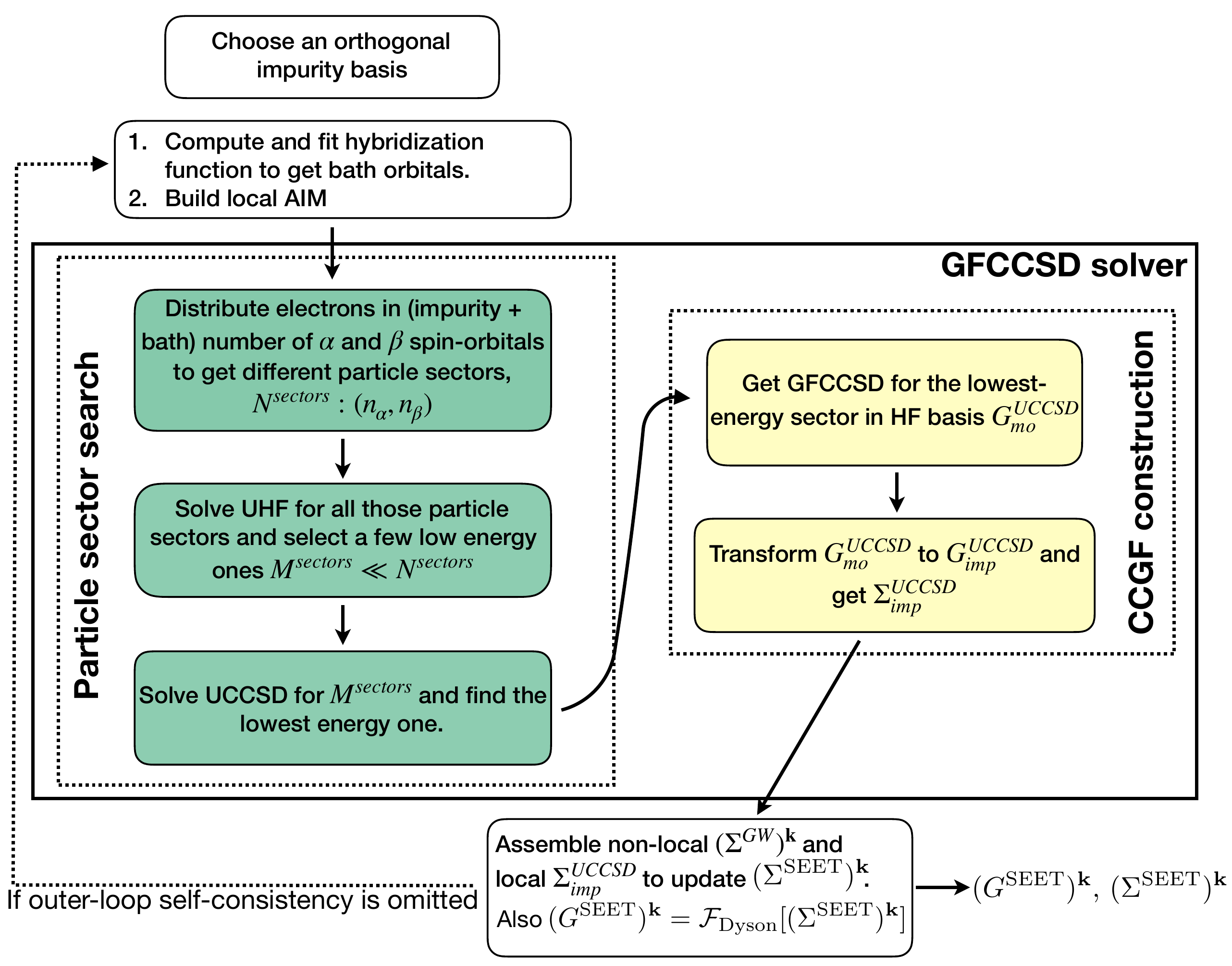}
    \caption{A schematic view of the inner loop and GFCCSD solver. Particle sector search and GFCCSD construction sections pertain to the solver, and the rest of the steps are general to the SEET scheme.}
    \label{fig:ccsd_solver}
\end{figure}

This search over different particle sectors is not only limited to closed shell particle sectors (containing even number of particles) but also includes all possible open-shell particle distributions (containing odd number of particles). Frequently, it is not straightforward to get unrestricted coupled cluster (UCC) solutions for each of the particle sectors. In those cases, the underlying unrestricted Hartree-Fock (UHF) method before the actual UCC step is unstable, and often converges to different local minima. To alleviate this problem we first carry out DIIS-based UHF steps and estimate the stability of the solution \cite{RudigerJCP97} after a certain number of iterations. If the solution vector is unstable, we rotate it to make it stable, and feed it to a Newton solver for final convergence. We must mention that converging to the desired root is quite essential, otherwise the ensuing SEET iterations show artifacts which are hard to control. Occasionally to facilitate the correct particle sector search, we use \textit{a priori} knowledge about the local spin moment of the impurity problem, which can restricts the search over sectors with different particle numbers. Examples of such restricted search will be discussed in Sec.~\ref{Sec:Results}.   

\subsubsection{GFCCSD construction}
CC is a many-body theory based on the exponential parametrization of the ket wave function $|\Psi \rangle = e^T |\Phi \rangle $ where $|\Phi\rangle$ is the reference mean-field wavefunction ($|\Phi_{\text{UHF}}\rangle$ in our cases). The bra wave function in CC is not an adjoint of the ket, because $e^T$ is not a unitary operator. The choice of the bra state is non-unique in CC, where we define $\langle \Psi| = \langle \Phi| (1+\Lambda) e^{-T}$ through the de-excitation operator $\Lambda$, thereby the bra and ket states are biorthonormal. A CC wave function is mapped to a corresponding Green's function \cite{Avijit19,NooijenIJQC1992,KowalskiJCP2014,ZhuPRB19} by using the Lehmann representation 
\begin{align}
G^{CC}_{pq} (\omega) = {} & \langle \Phi | (1+ \Lambda) \overline{a_p^\dagger} \frac{1}{\omega + \mu + \overline{H} - i\eta } \overline{a}_q| \Phi \rangle \nonumber \\ 
     {} & +  \langle \Phi | (1+ \Lambda) \overline{a}_p \frac{1}{\omega + \mu - \overline{H} + i\eta}  \overline{a_q^\dagger} | \Phi \rangle \label{Eq:ccgf2},
\end{align}
where, $\overline{a}_p = e^{-T} a_p e^T$, $\overline{a_p^\dagger} = e^{-T} a_p^\dagger e^T$ with $a^{\dag}_{p}$ ($a_{p}$) being creation (annihilation) operators to the single-particle state $p$, and $\overline{H} = e^{-T} H e^T - E_{gr}$. $E_{gr}$ is the UCCSD ground state energy. 
CC is equivalent to ED if no truncation is made in the rank of the cluster operator T. However, in a practical implementation, we approximate T as only singles and doubles, that is T = T$_1$ + T$_2$, which leads to a method that scales as $n^6$, where $n$ is the number of orbitals. In addition to solving the ground state problem, GFCCSD requires an additional step where we tridiagonalize $\overline{H}$ in the space of (N+1) and (N-1) electronic wave functions, using the Lanczos method. This particular step scales as $n^5$ for each of the elements $G^{UCCSD}_{pq}$, thus keeps the computational cost low.  Finally, the inversion of Eq. \ref{Eq:ccgf2}  is carried out using a continued fraction formula. Consequently, the computational scaling in our approach is independent of the size of frequency grid. 

Since GFCCSD is not written as a diagrammatic expansion of the Luttinger-Ward functional $\Phi$, GFCCSD is not a conserving approximation and therefore a non-causal self-energy is possible. Explicitly, it means that the imaginary part of the diagonal elements of the GFCCSD self-energy does not need to remain strictly negative. 

Note that CC calculations are usually carried out in a one-particle basis that may come from either UHF or unrestricted DFT calculations. Therefore G$^{UCCSD}$ that we first construct is in that basis. In a later step, we transform it from a given one-particle basis to the  original impurity basis.

\section{Results}\label{Sec:Results}
\subsection{Computational details}
Simulations were done at temperature $T\sim451$ K ($\beta = 700$ Ha$^{-1}$) for the MnO solid and $T\sim1053$ K ($\beta = 300$ Ha$^{-1}$) for the SrMnO$_3$ solid. We used the \emph{gth-dzvp-molopt-sr} basis~\cite{GTHBasis} with \emph{gth-pbe} pseudopotential~\cite{GTHPseudo} for all atoms in both the MnO  and the SrMnO$_3$ solids. 
In our $GW$ implementation~\cite{Iskakov20}, density fitting for four-index Coulomb integrals~\cite{Werner03,Ren12,Sun17} is utilized for the computational efficiency and memory reduction. 
Explicitly, the four-fermion Coulomb integrals were decomposed into a combination of even-tempered Gaussians for the Strontium atom and \emph{def2-svp-ri}~\cite{RI_auxbasis} for all the other atoms. 
The Coulomb integrals and non-interacting matrix elements were prepared by the open source \texttt{PySCF}~\cite{PySCF} package. 
\subsection{MnO solid}

\begin{figure}[htp]
\includegraphics[width=0.47\textwidth]{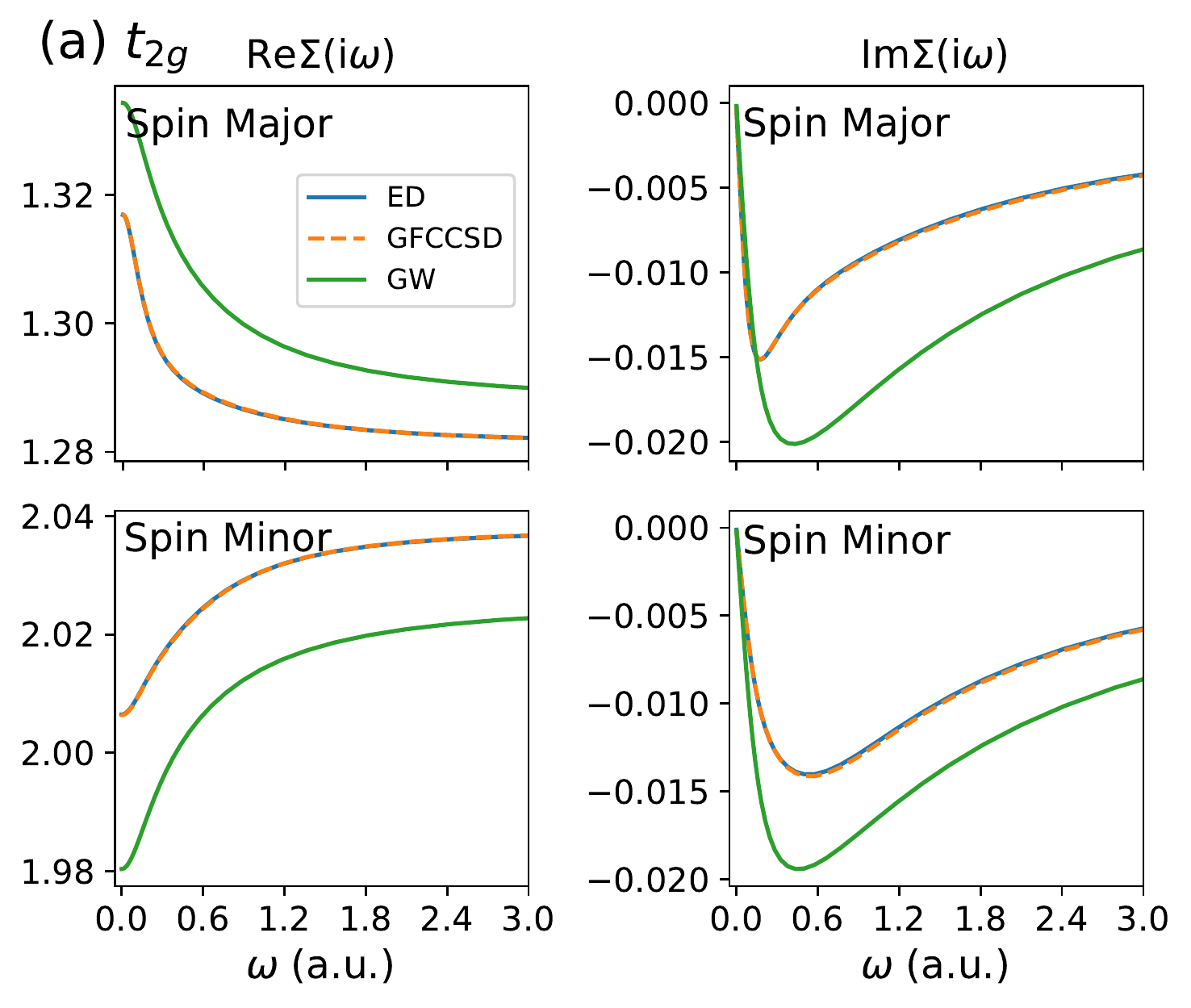}
\includegraphics[width=0.47\textwidth]{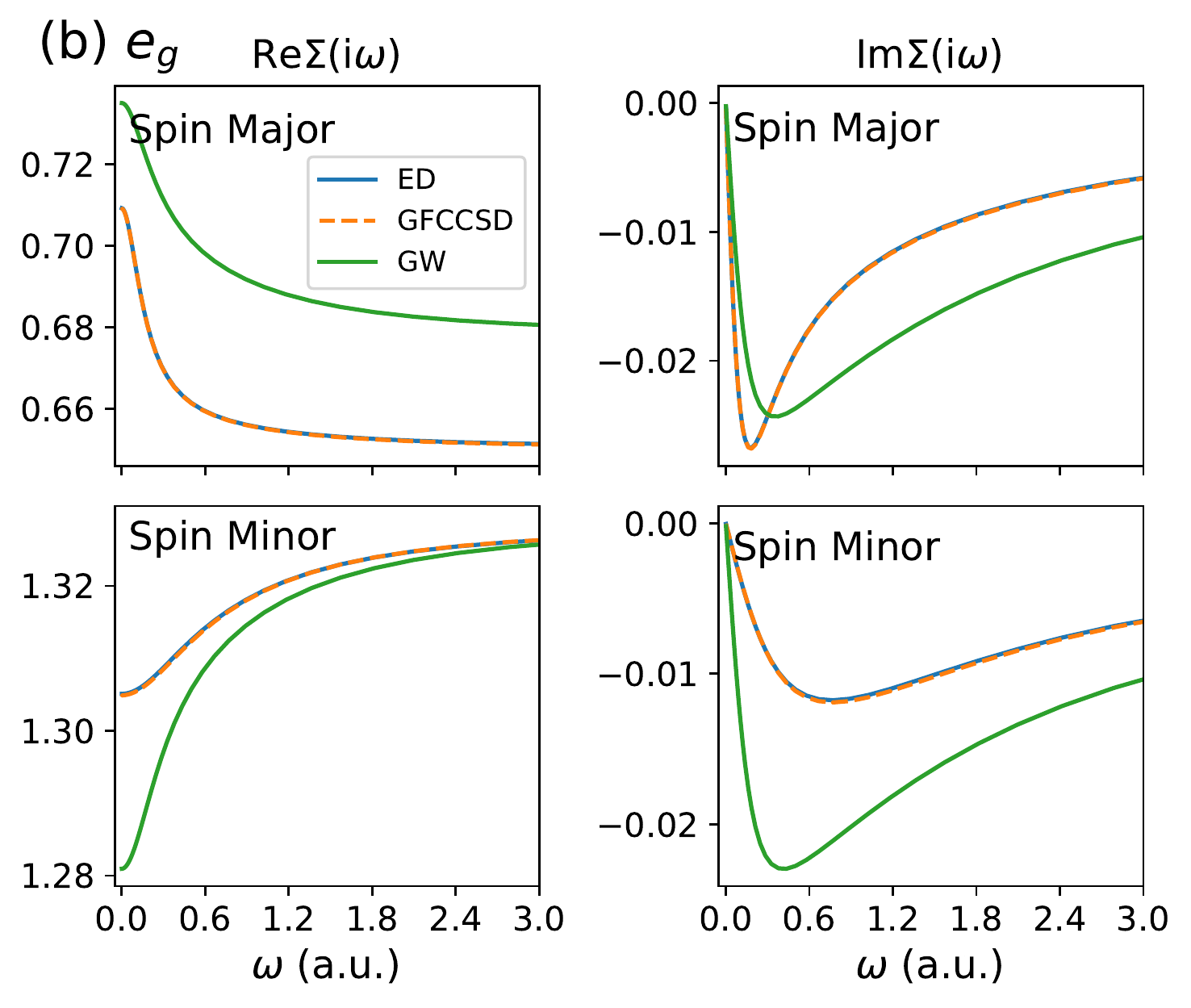}
\includegraphics[width=0.47\textwidth]{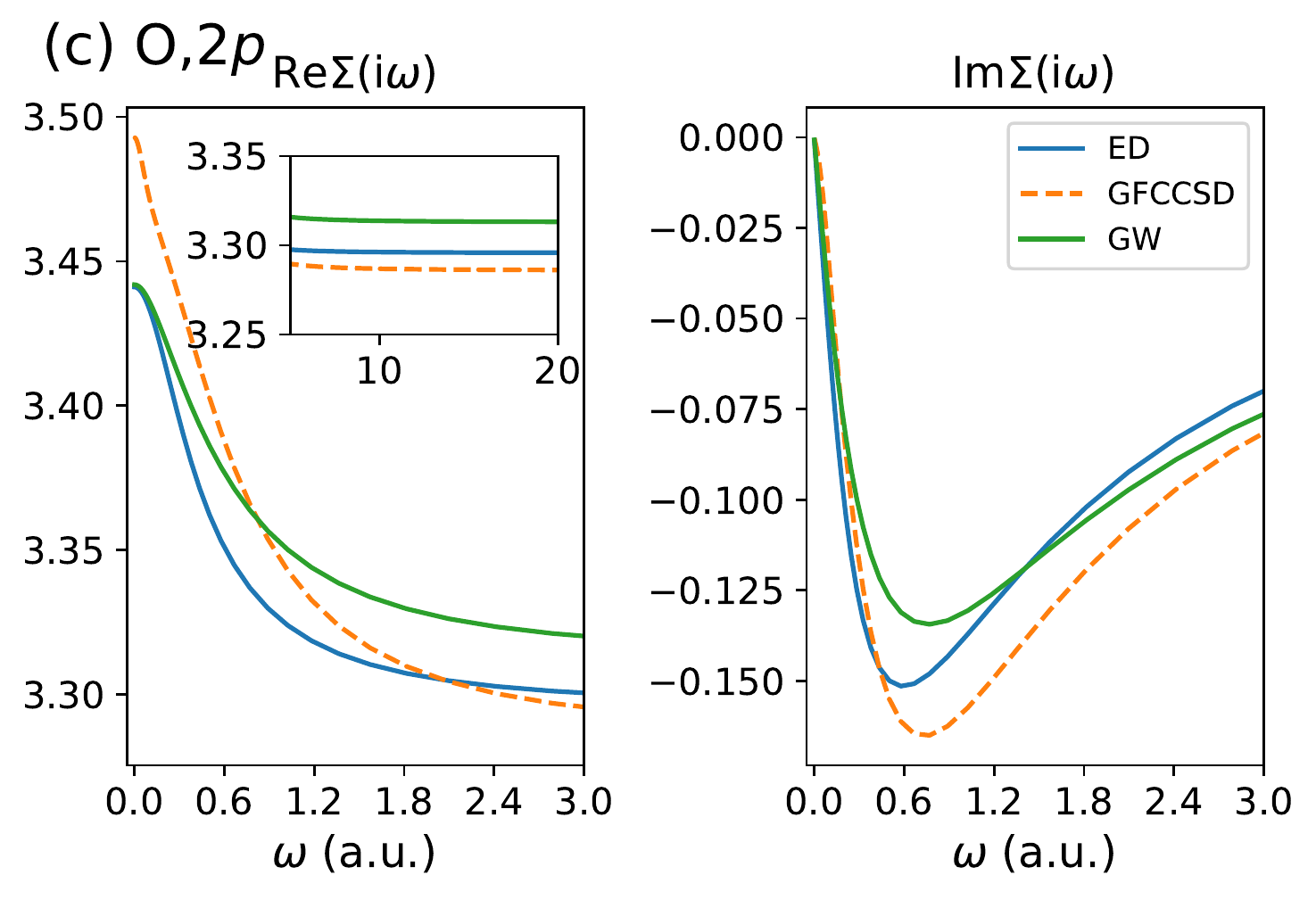}
\caption{The impurity self-energy $\Sigma$ comparison between ED and GFCCSD on imaginary frequency axis for the MnO crystal. $GW$ label is used to denote the double counting contribution coming from performing one iteration of $GW$ in the impurity orbital subset.}\label{fig:MnO_selfenergy_compare}
\end{figure}

\begin{figure*}[htp]
\includegraphics[width=0.47\textwidth]{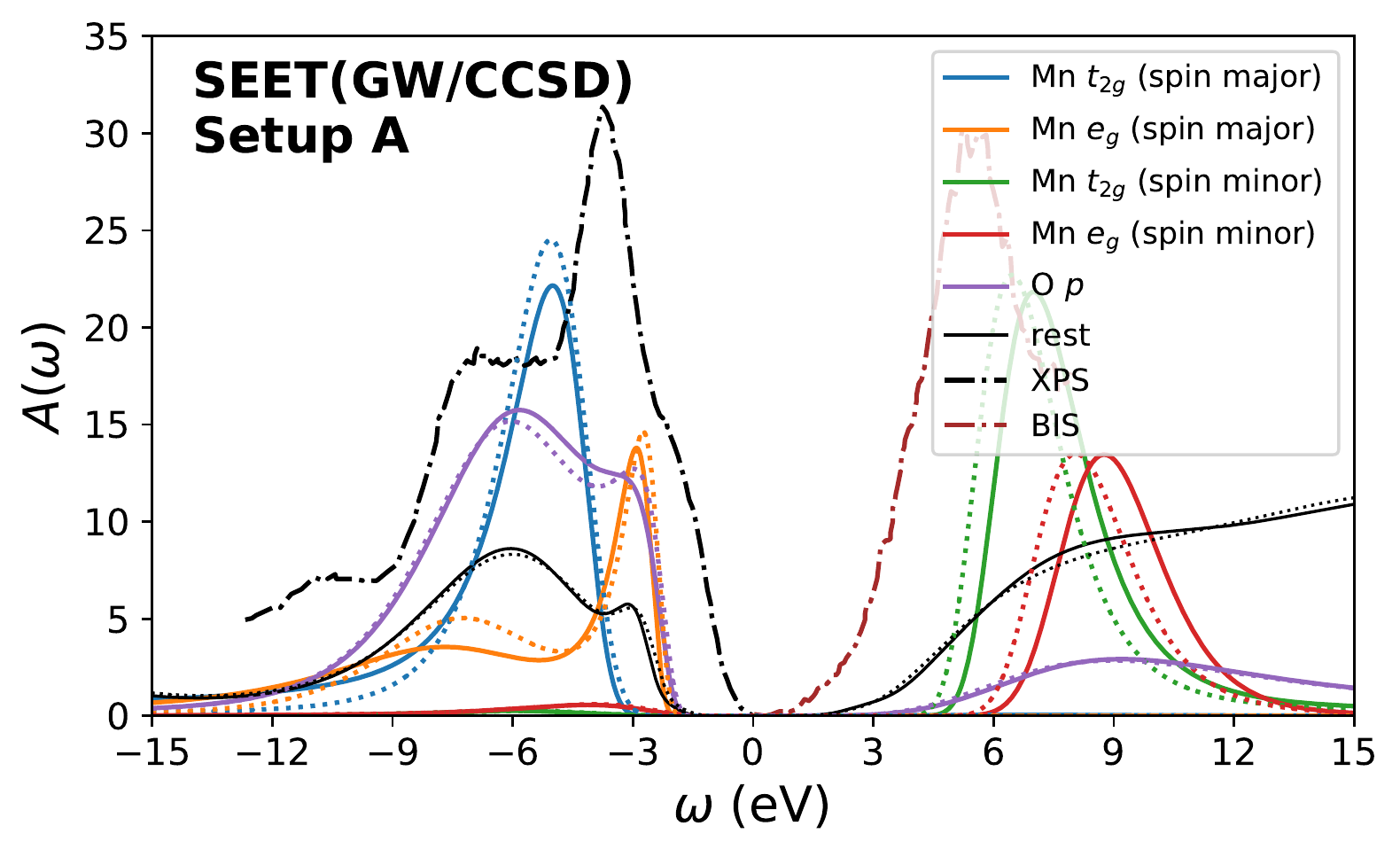}
\includegraphics[width=0.47\textwidth]{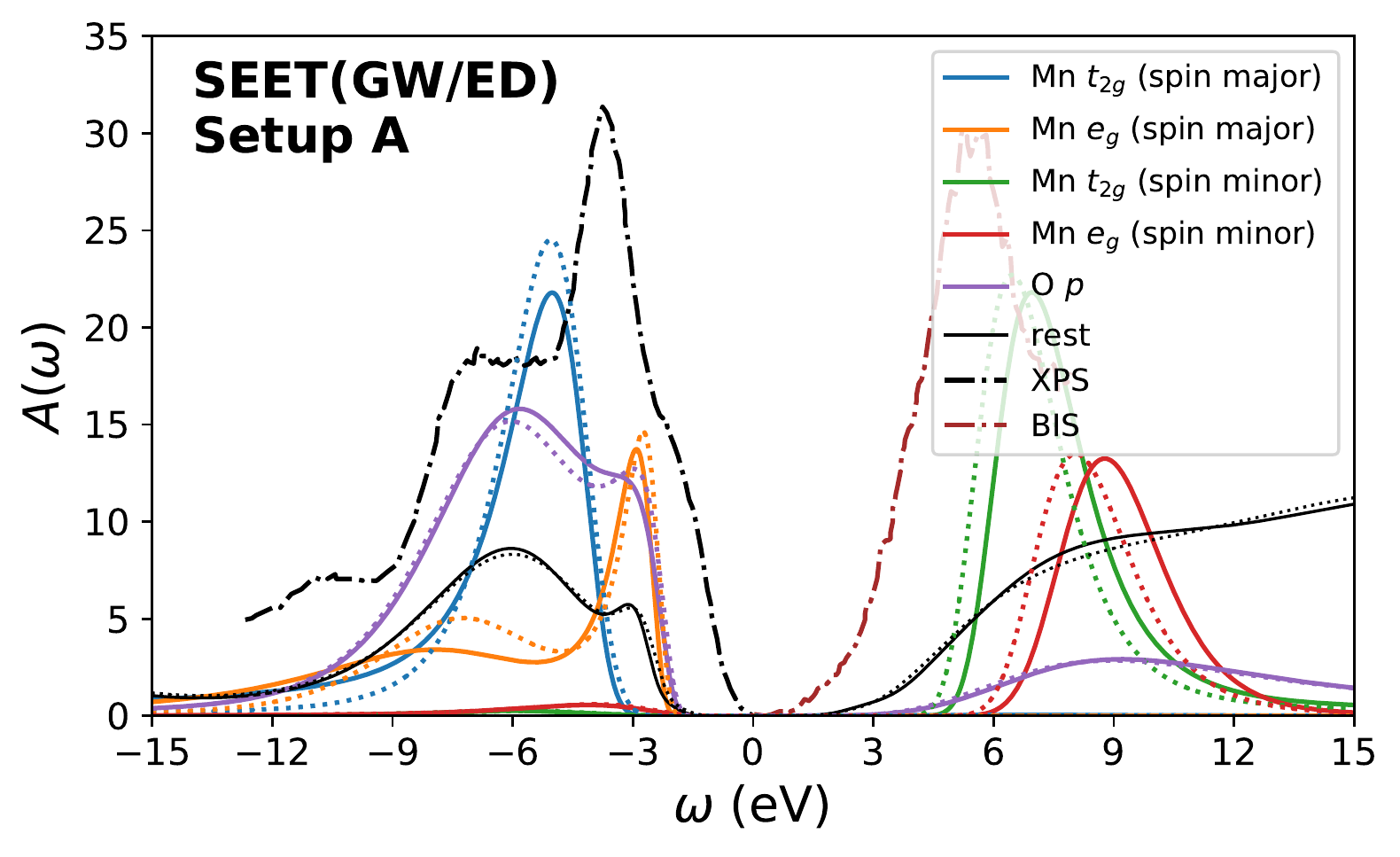}
\includegraphics[width=0.47\textwidth]{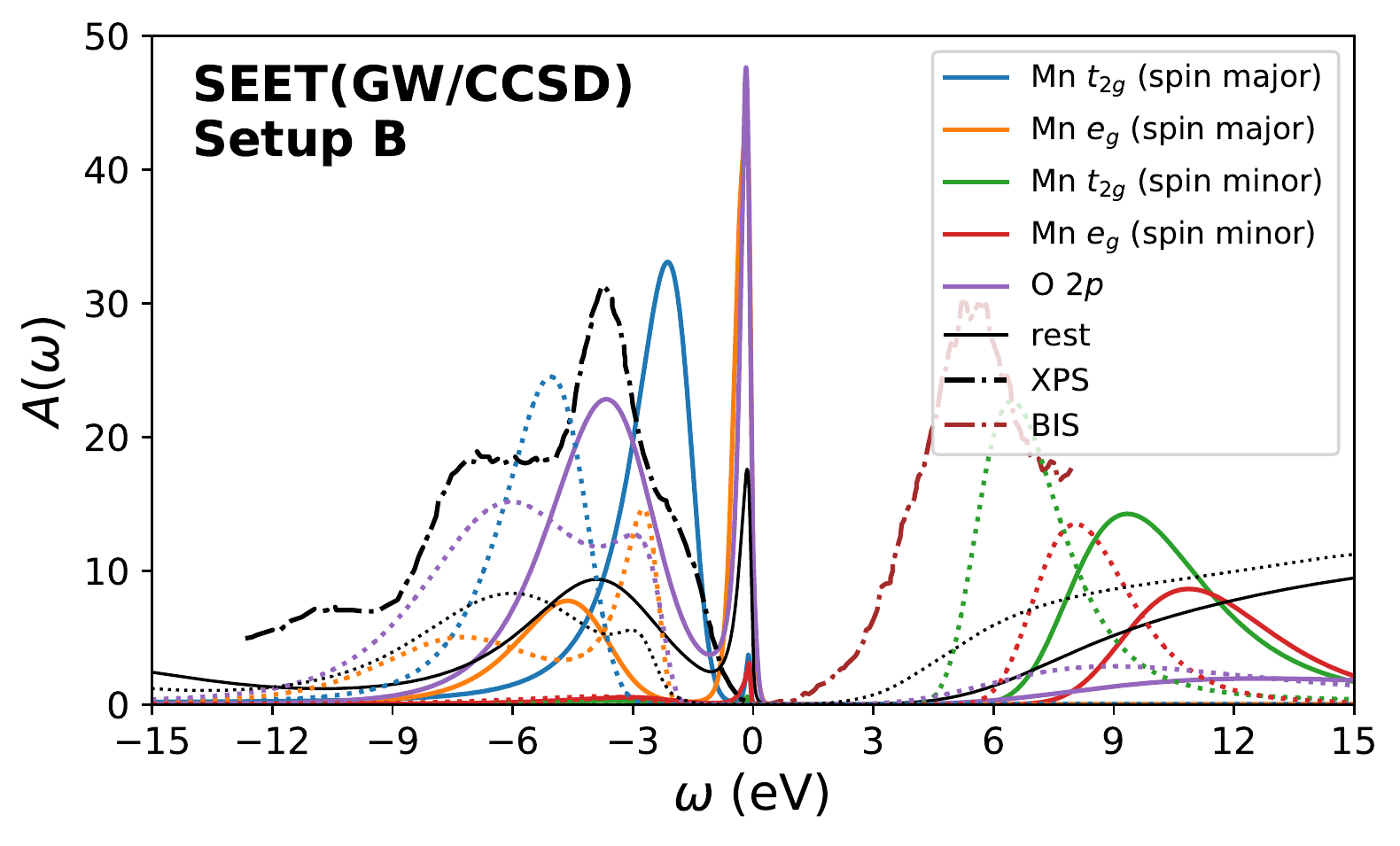}
\includegraphics[width=0.47\textwidth]{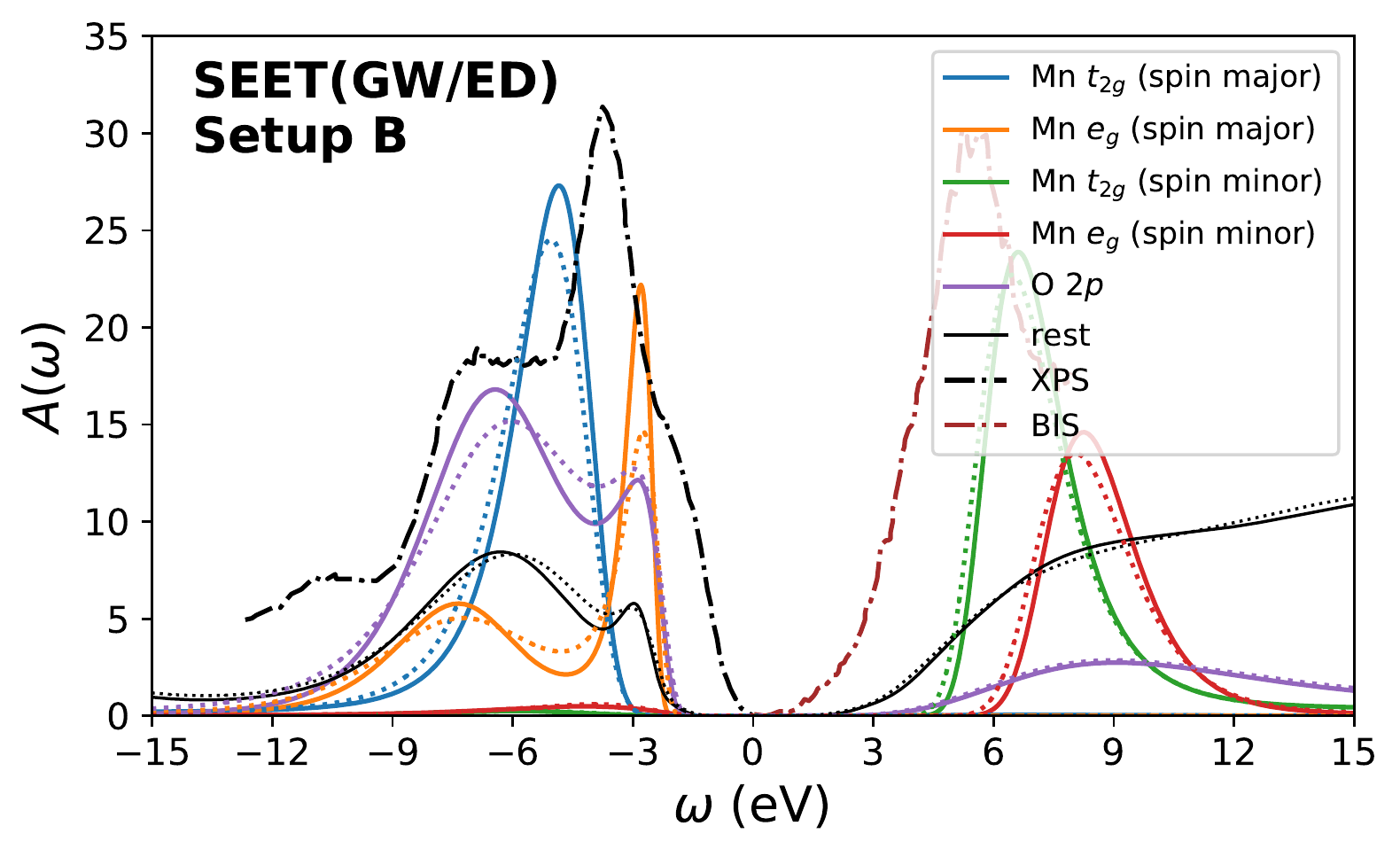}
\caption{Orbital-resolved local DOS for the MnO crystal from SEET($GW$/CCSD) and SEET($GW$/ED). The impurity choices from the first to second row correspond to (A) and (B) in Table \ref{tab:impurities_MnO}. Solid lines in both right and left panels are from SEET($GW$/CCSD) and SEET($GW$/ED) data, respectively. Dashed lines are photoemission data from Ref.~\cite{van91}. The dotted lines correspond to the orbitally resolved sc$GW$ calculation.}\label{fig:MnO_DOS_AB}
\end{figure*}

MnO is one of the prototypical strongly correlated systems with AFM ordering (The Ne\'{e}l temperature $\sim$ 120 K~\cite{Shull51}). 
Nominally, it has five singly occupied electrons in the Mn $3d$ shell. 
Due to its simple rocksalt structure, it has been an ideal testbed for strongly correlated methods. 
In order to simulate the AFM ordering, we double the primitive cell along [111] direction. 
The corresponding crystal supercell contains two manganese (Mn) and two oxygen (O) atoms in a rhombohedral unit cell.
Due to its charge-transfer nature, we consider not only Mn $3d$ but also O $2p$ orbitals in our choices of impurity problems. 
Correlations from different orbitals around Fermi surface ($E_{F}$) are disentangled by splitting Mn $3d$ and O $2p$ orbitals into several disjoint smaller impurity problems. 

\subsubsection{Comparison of impurity self-energies using ED and CCSD}\label{subsubsec:selfenergy_compare}

To assess how well GFCCSD works as a solver for realistic impurity problems created by splitting Mn $3d$ and O $2p$ orbitals into several smaller impurity problems, we compared the GFCCSD self-energies to the ones obtained from ED. These comparisons are important since they allow us to assess in a realistic context what to expect from GFCCSD.

In Fig.~\ref{fig:MnO_selfenergy_compare}, we plot the impurity self-energies on the Matsubara axis from ED and GFCCSD for MnO in the AFM phase. 
The number of bath orbitals are kept the same for the two solvers to avoid any discrepancy due to the bath discretization. 
In order to eliminate any cumulation or cancellation of errors due to the  self-consistency procedure, the data is extracted from the first iteration of SEET. 
In addition to self-energies from ED and GFCCSD, we also plot double counting contributions extracted from a single iteration of $GW$ within the impurity orbitals and denoted in Fig.~\ref{fig:MnO_selfenergy_compare} as $GW$. 
Since the impurity self-energies contain both static and dynamic parts, the high frequency limit of the real part of the self-energy decays to the static part of the self-energy. 

From the (a) and (b) panels of  Fig.~\ref{fig:MnO_selfenergy_compare}, we observe that for both Mn $t_{2g}$ and $e_{g}$ orbitals, GFCCSD reproduces the ED self-energy remarkably well in both low and high frequency regimes. 

The disagreement between GFCCSD and ED self-energies becomes clear when a non-perturbative treatment of O $2p$ is included, see Fig.~\ref{fig:MnO_selfenergy_compare} panel (c). 
Although the static part of the self-energy from GFCCSD and ED are close to each other (see the inset in Fig.~\ref{fig:MnO_selfenergy_compare} (c)), the dynamic counterpart is quite different for both the real and imaginary components. 
In the view of many-body theory, this implies that for the O $2p$ impurity, the GFCCSD approximation is only capturing well first-order Feynman diagrams. To recover the dynamic part of the O $2p$ impurity higher orders of self-energy diagrams are necessary and these require higher excitations in the CC theory. 

The manifestation of the GFCCSD's difficulty in handling the O $2p$ orbital impurities can be already observed at an earlier stage during the particle sector search where UCCSD is used to find an optimal number of particles for the impurity. 
In order to stabilize this search and the overall calculation, we constrained the particle sectors to the states that contain the same number of $\alpha$ and $\beta$ electrons. This constraint is determined and justified by the exact particle sectors found by ED. 
The difficulties and failure of UCCSD in the particle-sector search is already a sign of a problematic nature of GFCCSD for the impurity made out of  O $2p$ orbitals.

Through comparisons of the impurity self-energies from GFCCSD, ED and the $GW$ double counting counterpart, we found the O $2p$ orbitals to be the most strongly correlated ones while the Mn $3d$ shell is merely moderately correlated. 
These observations suggest that conventional impurity choices for MnO with the Mn $3d$ shell only are insufficient. 
Note that our impurity orbitals are symmetrized atomic orbitals (SAO)~\cite{Lowdin70_SAO} based on our choices of Bloch Gaussian basis which can only be viewed as approximations to the true physical atomic orbitals. 

\begin{table}[tbh]
\begin{ruledtabular}
\begin{tabular}{c|c|p{6cm}}
Name & Imp & Description \\
\hline
A & 2 & Mn $t_{2g}$; Mn $e_{g}$ for a single Mn atom\\
B & 1 & O $2p$ \\
C & 1 & all $3d$ orbitals for both Mn atoms in the unit cell
\\
\end{tabular}
\end{ruledtabular}
\caption{Different choices of impurities for the MnO solid. Imp denotes the number of distinct, disjoint impurity problems.}
\label{tab:impurities_MnO}
\end{table}

\subsubsection{Local density of states}

Here, we consider the effect of using the GFCCSD solver on the local density of states obtained in the SEET($GW$/CCSD) procedure. Note that to assess the accuracy of this method, we compare the obtained local density of states to the one from the SEET($GW$/ED) data. 
In Fig.~\ref{fig:MnO_DOS_AB}, we display the orbital-resolved local density of states (DOS) of the MnO solid from SEET($GW$/CCSD) and SEET($GW$/ED) for different impurity setups (for details see Table.~\ref{tab:impurities_MnO}), analytically continued from the imaginary to the real frequency axis by Maxent~\cite{Jarrell96,Levy17}.
The x-ray photoemission (XPS) and bremsstrahlung isochromat spectroscopy (BIS) data are also included and shown for comparison~\cite{van91}. 
Based on our previous work on MnO~\cite{Iskakov20}, the outer-loop self-consistency of SEET is not performed since we observed that it did not change the results.
In the computational supercell there are two identical MnO units with opposite local moments on Mn atoms. The impurities in setups A and B created for these two primitive units contained in the supercell are disjoined. Only one of such disjoined impurities has to be calculated, the other one is recovered considering symmetry between spins. This means that the many-body correlation effects that  describe interactions between these impurities in the supercell are described at the GW level while the impurity orbitals themselves are treated by a higher level methods (here either ED or GFCCSD). 


Since for impurity choices denoted as setup A with two disjoint impurities $t_{2g}$ and $e_{g}$ on each of the  Mn atoms no major deviations in Mn $e_{g}$ and  $t_{2g}$  were observed in the first iteration (see panel (a) and (b) of Fig.~\ref{fig:MnO_selfenergy_compare}) the self-consistent DOS from SEET($GW$/CCSD) and SEET($GW$/ED) are almost indistinguishable. 
As shown in the top row of Fig.~\ref{fig:MnO_DOS_AB}, GFCCSD successfully captures the inner local correlations within the Mn $3d$ shell. Here, the presence of correlations beyond $GW$ is responsible for renormalizing the valence Mn $t_{2g}$ and  $e_{g}$ bands and pushing the conduction Mn $t_{2g}$ and $e_{g}$ bands a bit further away from $E_{F}$. 

Next, as illustrated in the bottom row of Fig.~\ref{fig:MnO_DOS_AB}, we included a non-perturbative treatment of O $2p$ orbitals (setup B). Unlike the case in impurity setup A, the DOS from SEET($GW$/CCSD) is dramatically different from the DOS calculated using SEET($GW$/ED). 
Even though the orbital ordering remains mostly unchanged, significantly different chemical potential shifts are found in SEET($GW$/ED) and SEET($GW$/CCSD).
In  SEET($GW$/CCSD), the description of many-body effects present in the impurity orbitals at the GFCCSD level is responsible for the enhancement of the satellite peak for both O $2p$ and Mn $e_{g}$ in spin-major channel.

As we show in Sec.~\ref{subsubsec:selfenergy_compare}, O p orbitals show stronger correlation effects and can not be fully resolved by CCSD level of GFCC. The corresponding error could later  cumulate during the self-consistency of SEET($GW$/CCSD) and make the final results uncontrolable. 


\begin{figure}[htp]
\includegraphics[width=0.47\textwidth]{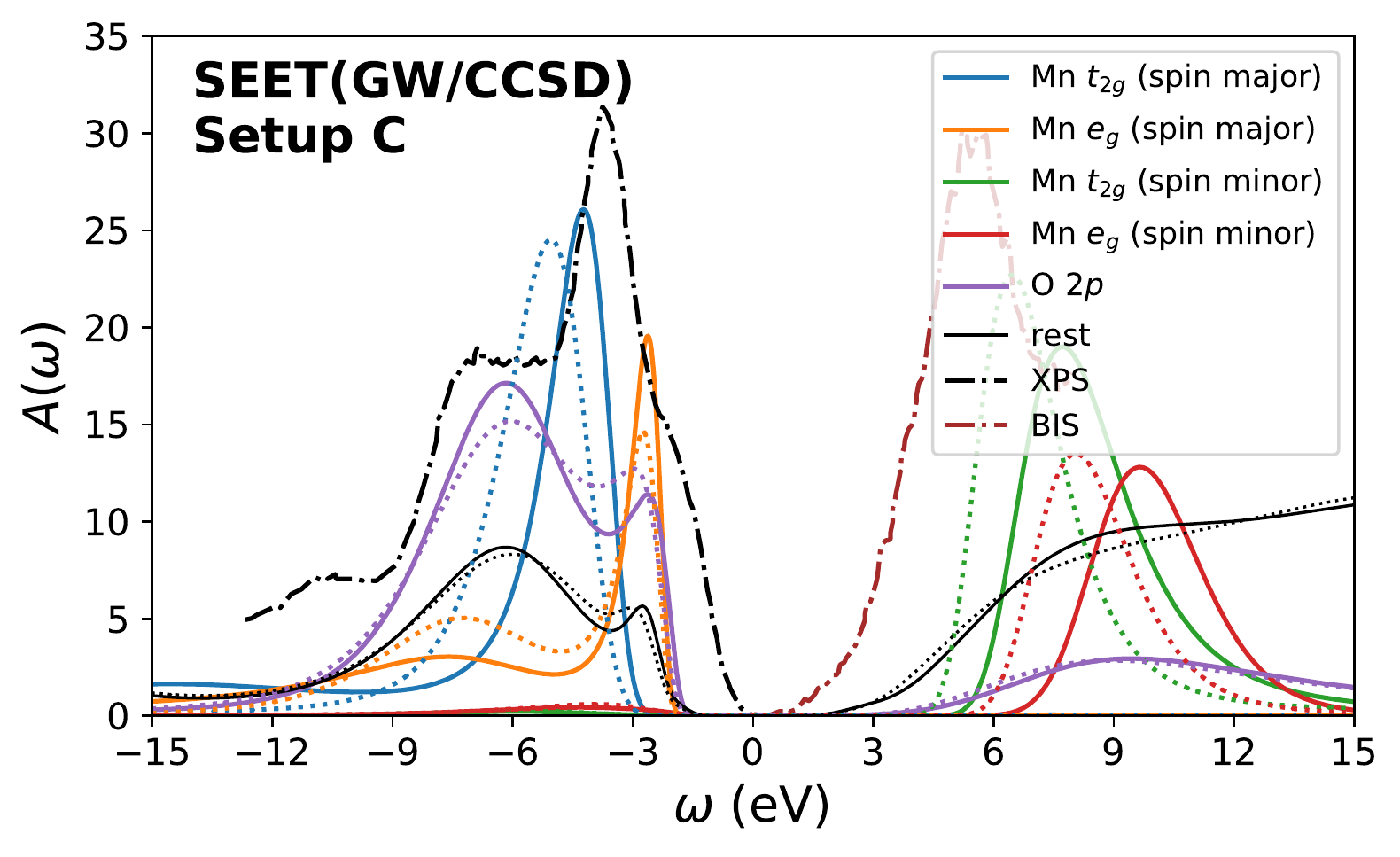}
\caption{Orbital-resolved local DOS of MnO from SEET($GW$/CCSD) for impurity setup C (see Table.~\ref{tab:impurities_MnO}). Solid lines are from SEET($GW$/CCSD) data, dashed line are photoemission data from ref~\cite{van91} and the dotted lines correspond to the orbitally resolved sc$GW$ calculation. }\label{fig:MnO_DOS_C}
\end{figure}

The strength of GFCCSD lies in the ability to treat larger impurity problems than the ones treatable by ED.
In setup C, we combine the entire $3d$ shell from both Mn atoms into one big impurity. The size of this impurity is far beyond the capability of ED or any other state-of-art impurity solver such as continuous-time quantum Monte Carlo (CT-QMC)~\cite{gull11_CTQMC}. 
The corresponding DOS (Fig.~\ref{fig:MnO_DOS_C}), however, shows no qualitative difference compared to the DOS from the impurity setup A. Only a slightly more renormalization to the satellite peaks of Mn $e_{g}$ in the spin-major channel and O $2p$ is observed. 
This implies that the inter-correlations between Mn $t_{2g}$ and $e_{g}$ and inter-correlations between two Mn sites are already well-captured by sc$GW$ and the corrections from the GFCCSD are not changing the spectrum.

Considering the fact that O $2p$ orbitals seem to be the most correlated ones in MnO, it is temping to combine O $2p$ orbitals with Mn $3d$ into one big impurity. 
Unfortunately, although it is computationally feasible, we found that SEET($GW$/CCSD) fails to converge in a  self-consistent manner for such a challenging impurity problem. This may be due to either (i) the instability of the particle sector search, and/or (ii) artificial spin orbital symmetry breaking  present in UCCSD. 
Both (i) and (ii) are consequences of the non-exactness of UCCSD and neither of them appears in SEET($GW$/ED).

Recent studies~\cite{Zhu20_HF_CCSD,zhu2020ab} have also utilized GFCCSD as the impurity solver in the DMFT-type embedding scheme where realistic impurity problems are constructed based on HF~\cite{Zhu20_HF_CCSD} and $G_{0}W_{0}$~\cite{zhu2020ab}. 
In these studies, the strength of GFCCSD is manifested by its ability to deal with an impurity problems which encompasses all the orbitals inside a unit cell. 
In our work, we take a somewhat different approach and we choose to include in the impurity problem only the orbitals that are significantly correlated leading to much smaller impurity problems.
While many steps in SEET($GW$/CCSD) and full cell $GW$+DMFT are similar, there is a number of differences that should be considered when comparing the results from these methods.

Intrinsic differences between SEET($GW$/CCSD) and full cell $GW$+DMFT arise due to (i) particle number present in the impurities and/or (ii) a different self-consistency condition. 
In the search for a particle number present in the impurity, in SEET($GW$/CCSD), we use UCCSD while in $GW$+DMFT reported in Ref.~\cite{Zhu20_HF_CCSD} the particle number presented in the impurity is determined by HF with fixed chemical potential. This is equivalent to using HF to find the ground state among the entire Fock space. We found, especially for impurities with strong correlations, that such a procedure could favor a choice of incorrect sectors and therefore prevents the self-consistency loop from converging (see Sec.~\ref{Sec:SrMnO3} and Table.~\ref{tab:sectors_order_p_pi} for more details).  Instead, in SEET($GW$/CCSD) the particle searching procedure based on UCCSD turns out to be more stable, although we still expect it to fail once correlations become too strong and the UCCSD approximation breaks down (see Sec.~\ref{Sec:SrMnO3} and Table.~\ref{tab:sectors_order_t2g}). 

Note that the problems arising during the particle number search in the impurity are universal and must be present for all the approximate wavefunction-based impurity solvers.  These problems appear since  different approximate methods can yield different orderings of ground state energies coming from different particle sectors.

As for the case (ii), the self-consistent condition is different in SEET when compared to the DMFT-type embedding frameworks. 
In SEET, the non-local $GW$ self-energy to the local correlated orbital is explicitly included in the SEET self-consistency condition (see Eqn. 25 in Ref.~\cite{Iskakov20}) and is not included in the hybridization function, in contrast to the procedure done in Refs.~\cite{Zhu20_HF_CCSD,zhu2020ab}. 

In addition, a further difference present in the full cell $GW$+DMFT implementation~\cite{zhu2020ab} relies on  the use of one-shot $G_{0}W_{0}$ rather than sc$GW$ (in our case). 
Empirically it is believed that $G_{0}W_{0}$ generally outperforms sc$GW$ due to cancellation of errors~\cite{Shishkin07,Kutepov17} while on the other hand severe dependency on the mean-field reference may occur especially when electronic correlations are strong which makes the procedure less \emph{ab initio}. 
Lastly, we want to emphasize that the choices of orthogonal basis for the impurity problem are also different in SEET and the full cell $GW$+DMFT. The orthogonal orbitals chosen for embedding in Refs.~\cite{Zhu20_HF_CCSD,zhu2020ab} are crystalline intrinsic atomic orbitals (IAOs) for the valence region and projected atomic orbitals (PAOs)~\cite{Knizia13} for the rest, while in SEET we use SAO basis~\cite{Lowdin70_SAO}. 
Consequently, a significantly different impurity Hamiltonian is obtained even though bare Coulomb interactions are used in both SEET and the full cell $GW$+DMFT implementation~\cite{zhu2020ab}. 

\subsection{SrMnO$_{3}$}\label{Sec:SrMnO3}
\begin{figure*}[htp]
\includegraphics[width=0.47\textwidth]{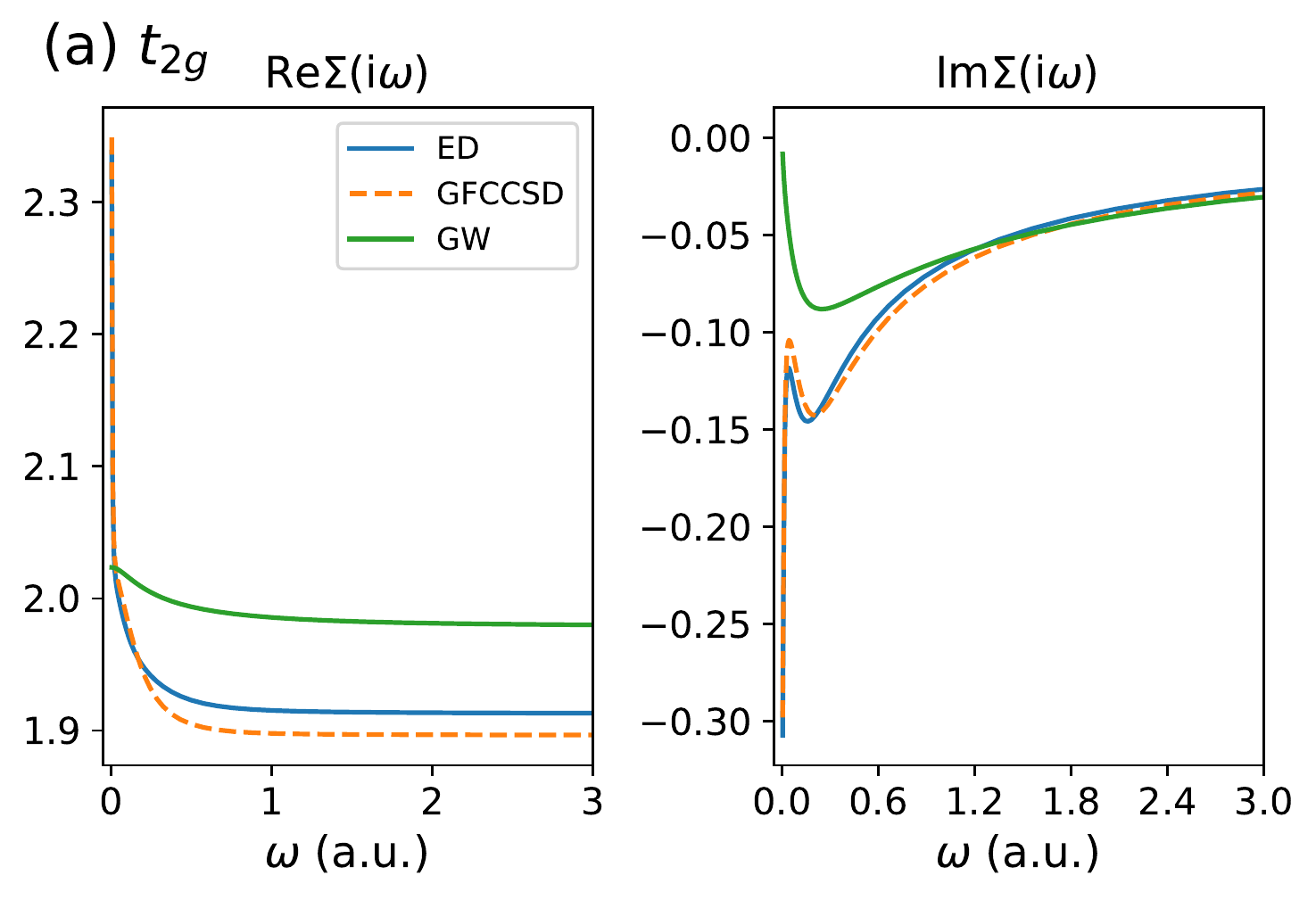}
\includegraphics[width=0.47\textwidth]{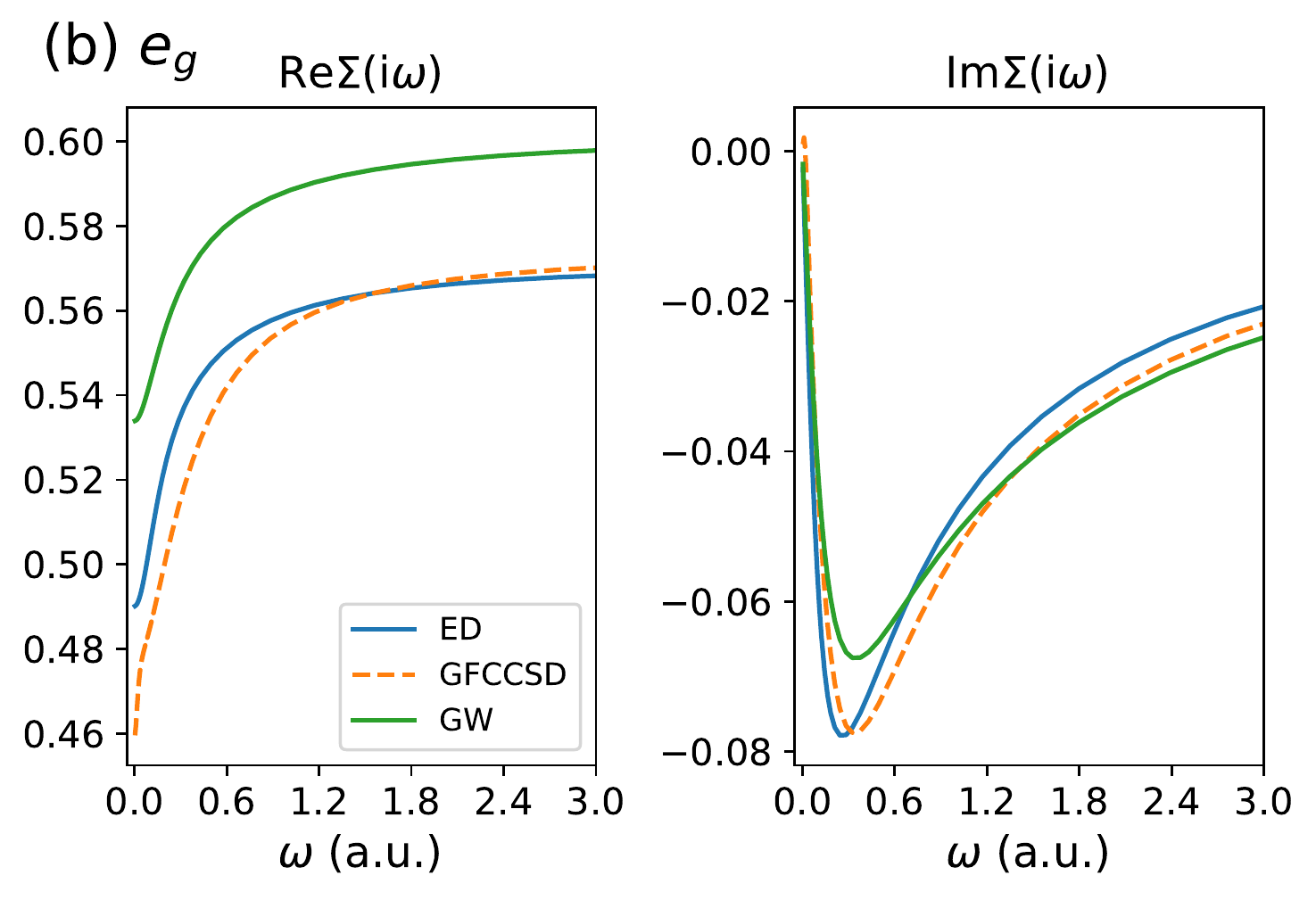}
\includegraphics[width=0.47\textwidth]{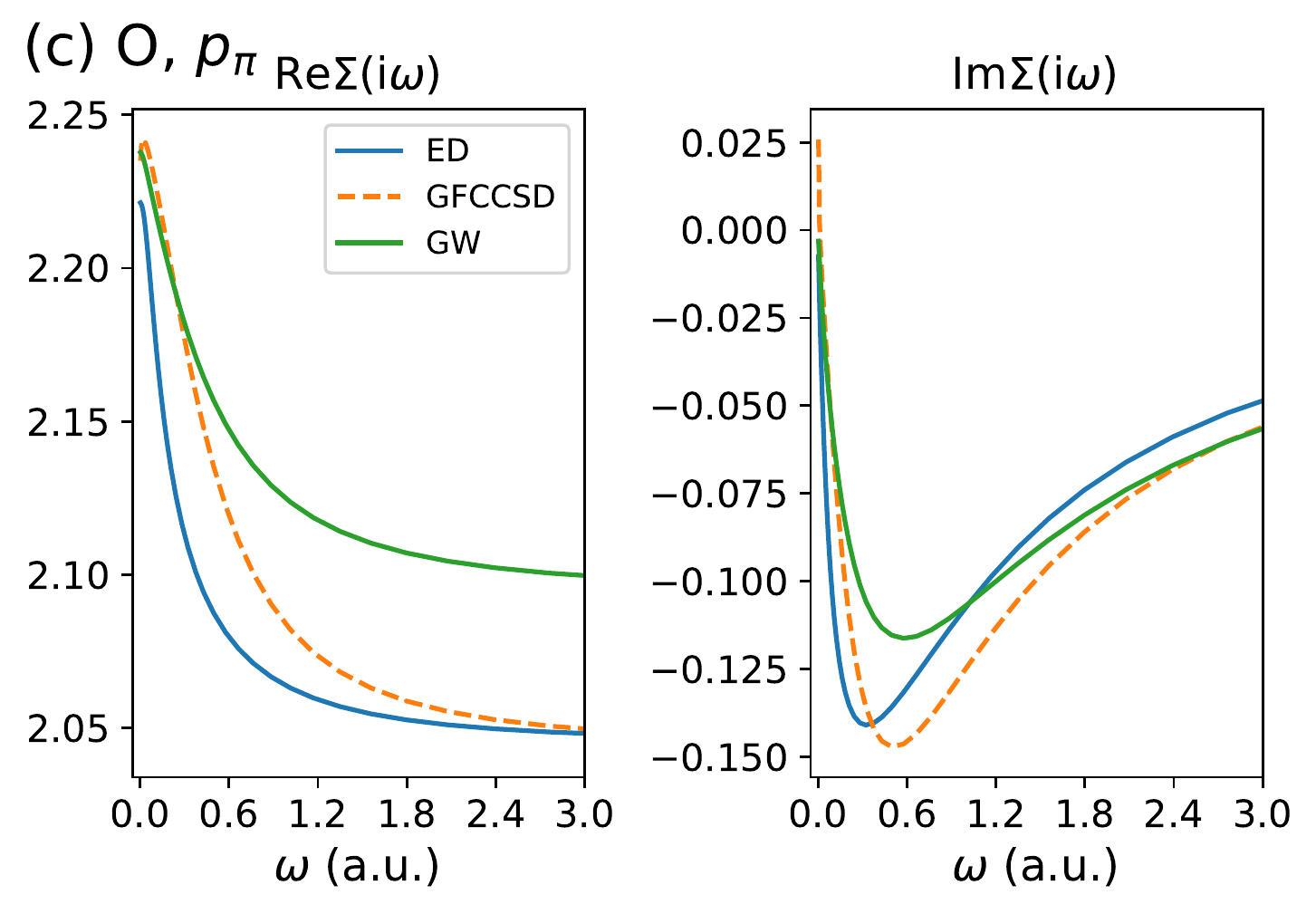}
\includegraphics[width=0.47\textwidth]{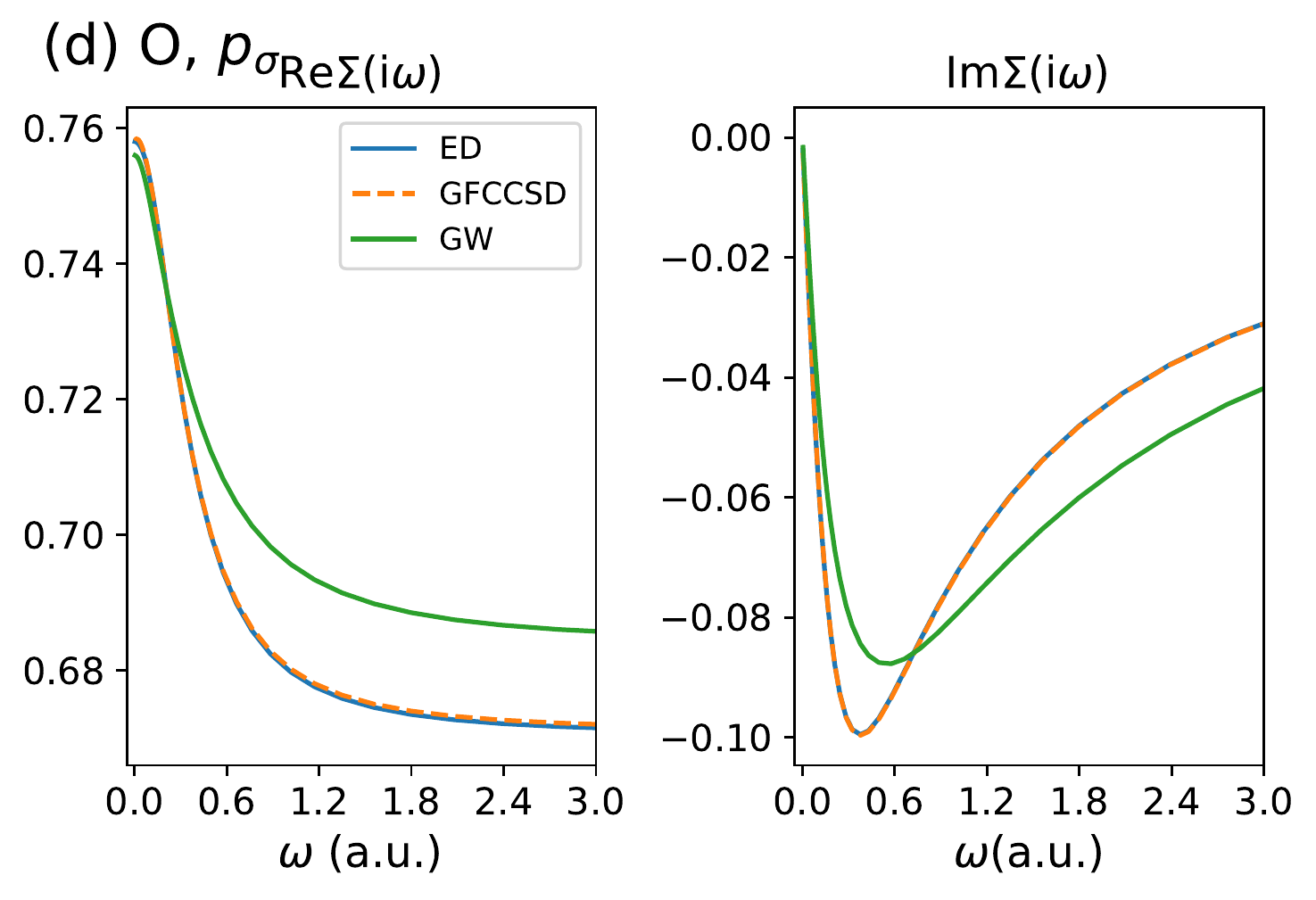}
\caption{
The impurity self-energy $\Sigma$ comparison between ED and GFCCSD solvers on the imaginary frequency axis for the SrMnO$_{3}$ crystal. $GW$ label is used to denote the double counting contribution coming from performing one iteration of $GW$ in the impurity orbital subset.}\label{fig:SrMnO3_selfenergy_compare}
\end{figure*}

\begin{figure*}[htp]
\includegraphics[width=0.47\textwidth]{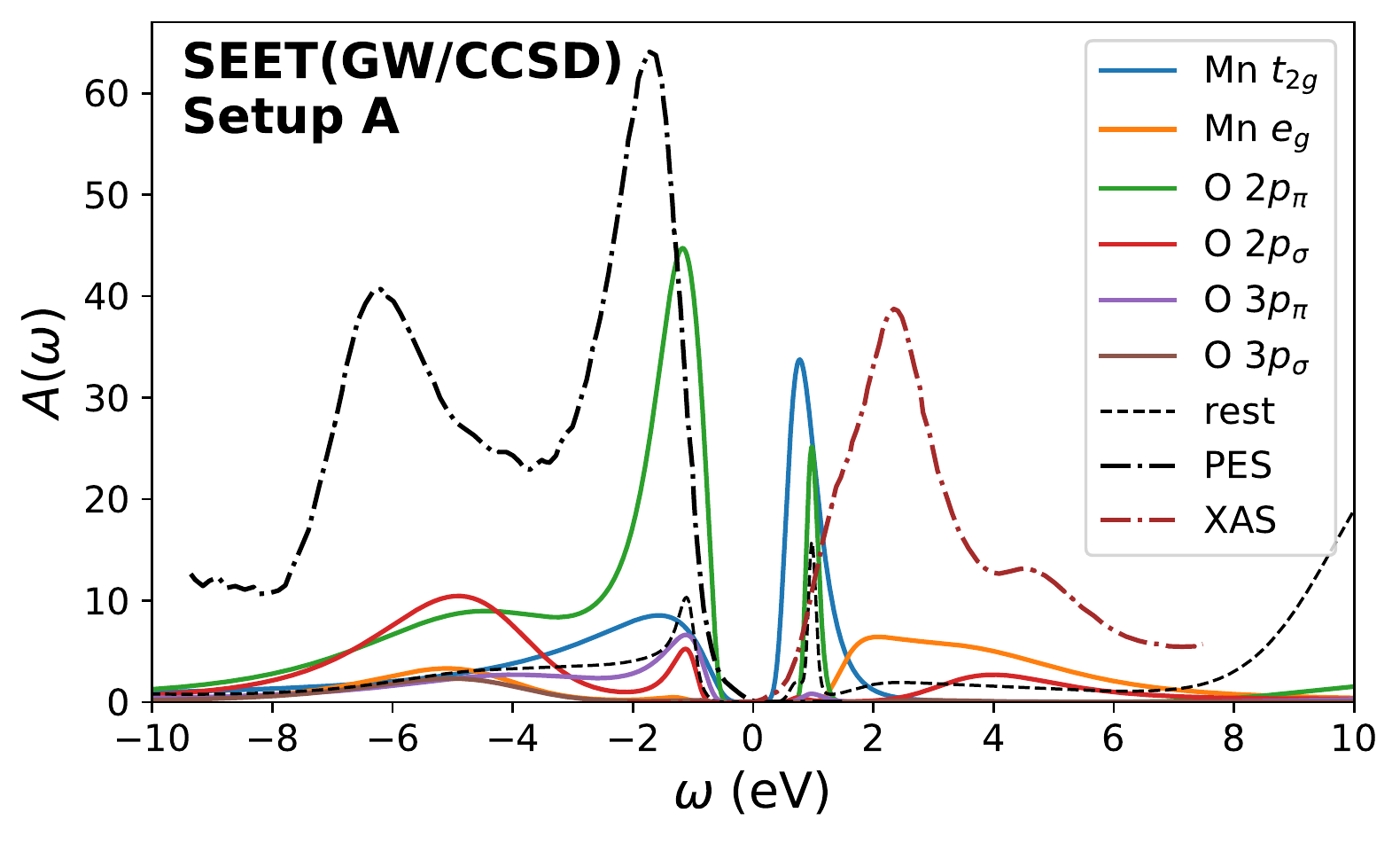}
\includegraphics[width=0.47\textwidth]{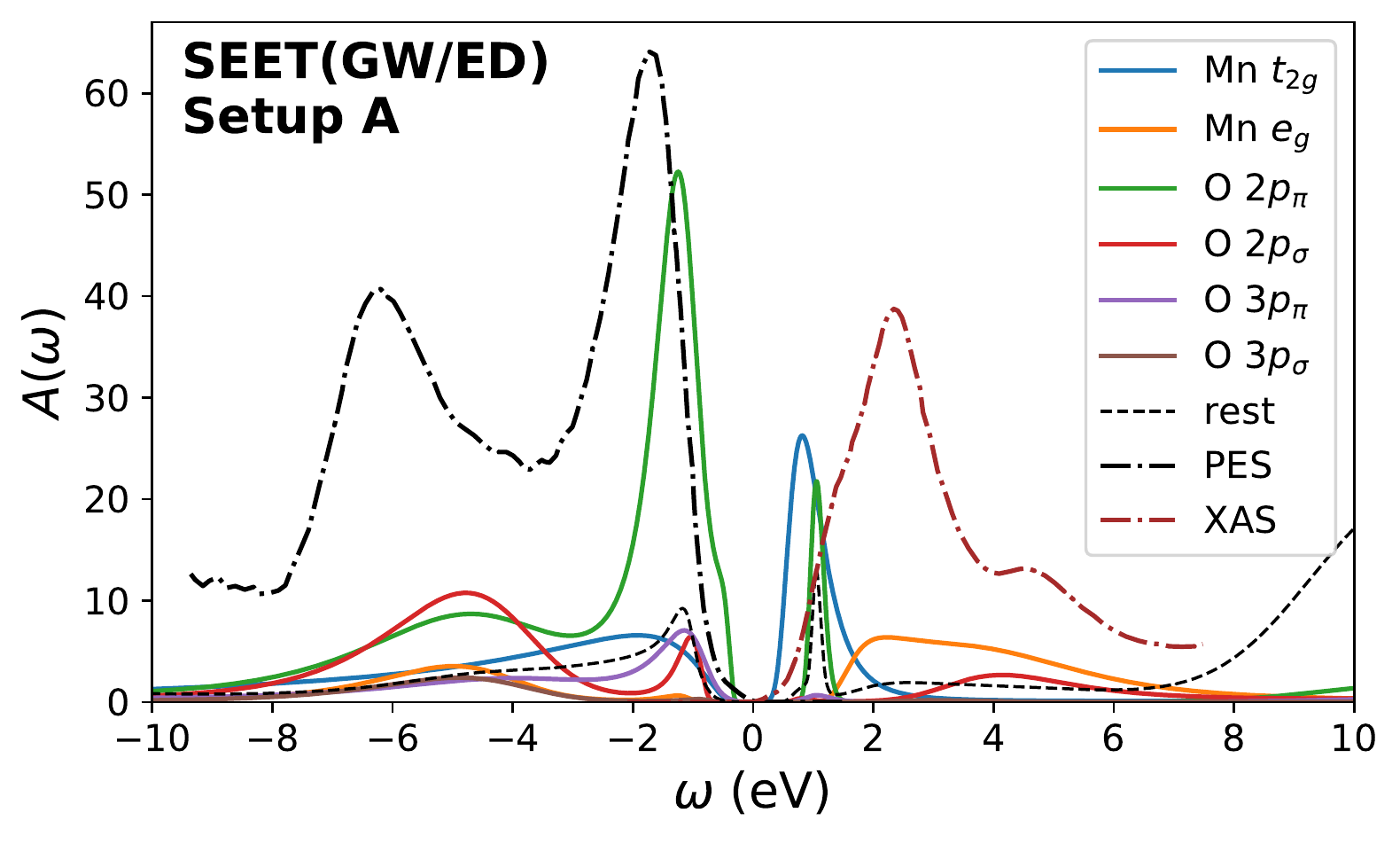}
\includegraphics[width=0.47\textwidth]{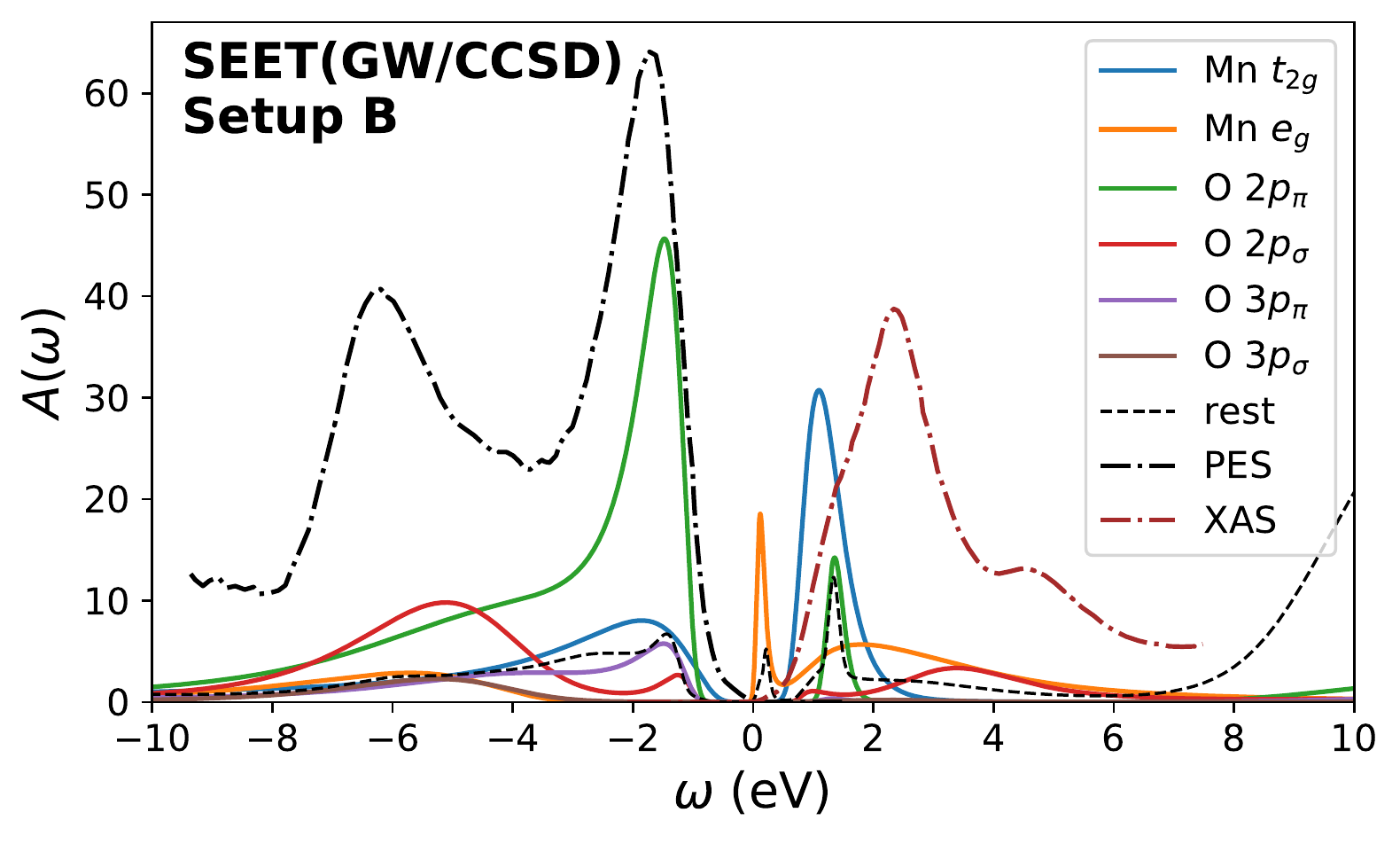}
\includegraphics[width=0.47\textwidth]{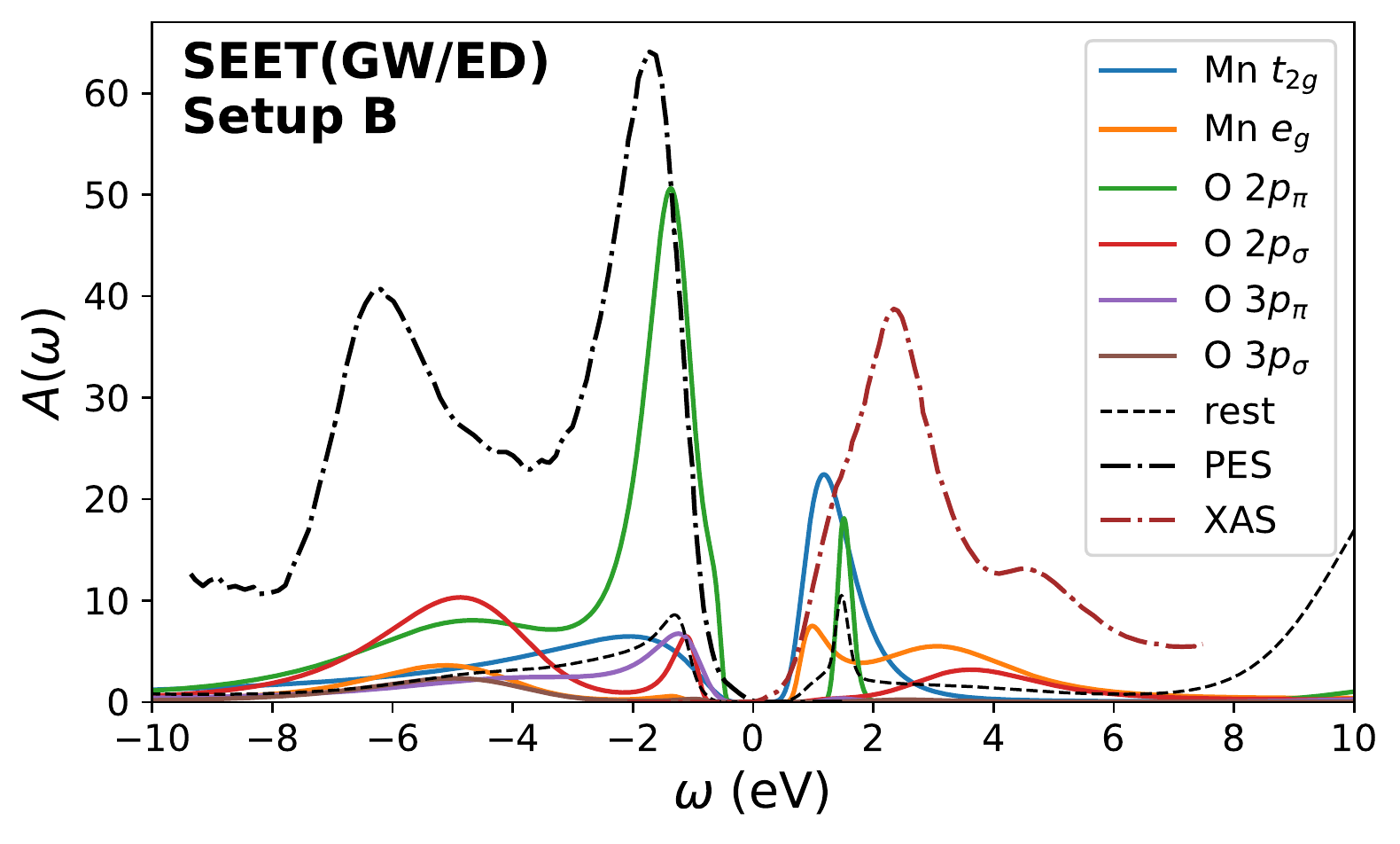}
\includegraphics[width=0.47\textwidth]{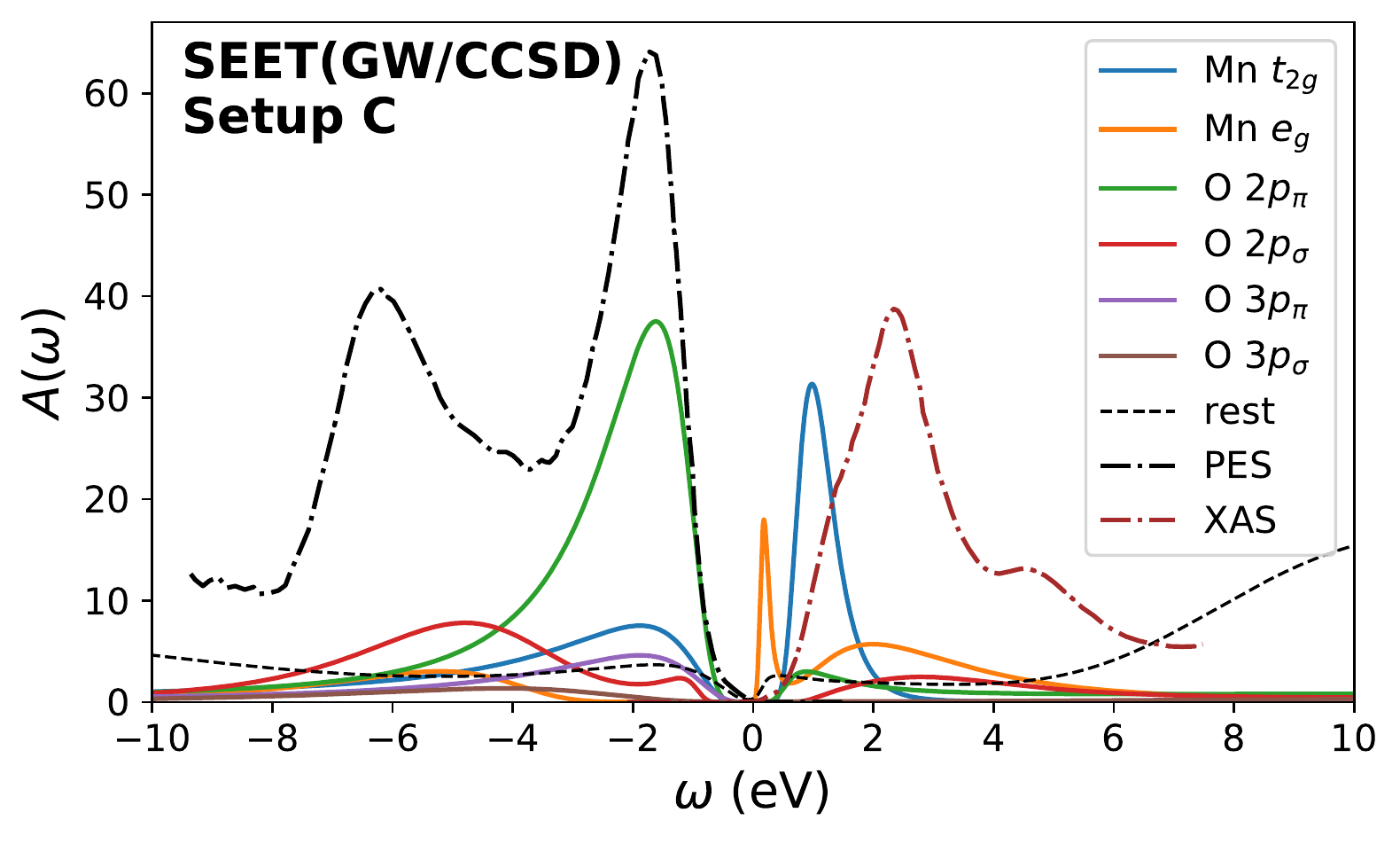}
\includegraphics[width=0.47\textwidth]{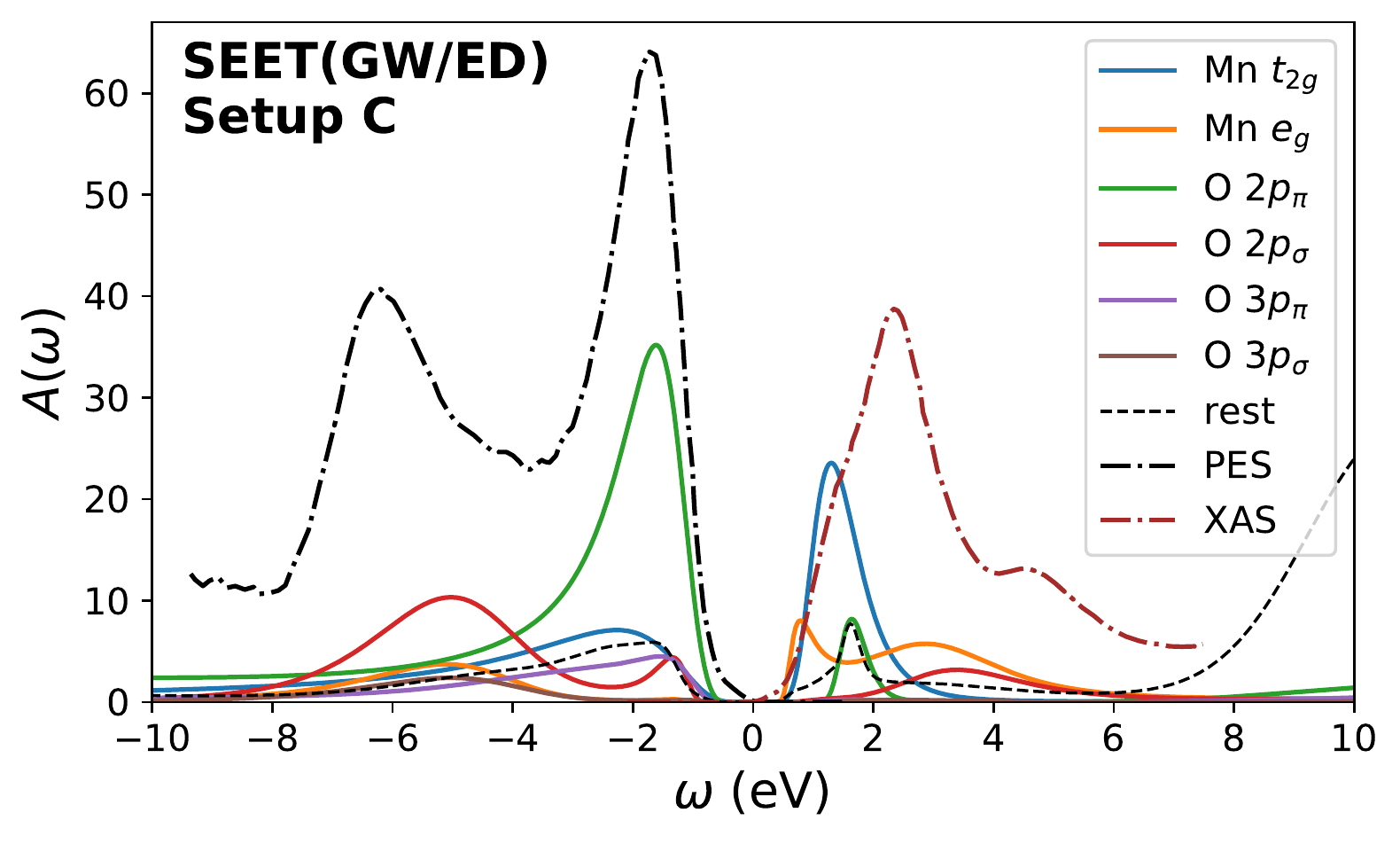}
\caption{Orbital-resolved local DOS for the SrMnO$_{3}$ crystal for  the  MnO  crystal  from  SEET($GW$/CCSD)  and  SEET($GW$/ED). The impurity choices from the first to third row correspond to (A), (B), and (C) in Table \ref{tab:impurities_SrMnO3}. Dashed line are photoemission data from Ref.~\cite{SrMnFeO3_expt_Kim10}.
}\label{fig:SrMnO3_PDOS}
\end{figure*}

SrMnO$_{3}$ is reported to be a cubic perovskite with $G$-type AFM ordering at Mn atoms with Ne\'{e}l temperature $\sim$ 233$-$260 K~\cite{Negas70,Takeda74}. Unlike MnO, it has nominally three singly occupied electrons on Mn $t_{2g}$ orbitals. 
Several theoretical studies for AFM phase of  SrMnO$_{3}$ were performed in Refs.~\cite{Rune06,Dang14,GW_EDMFT_PRM_Philipp20}. However, photoemission experiments~\cite{Abbate92,Saitoh95,Kang08,SrMnFeO3_expt_Kim10} are usually conducted at room temperature corresponding to the PM phase. 
In this work, we focus on the high-temperature PM insulating phase where sc$GW$ qualitatively predicts an incorrect metallic phase~\cite{Yeh20}. 
In addition, both LDA+DMFT~\cite{Dang14,Chen14,Bauernfeind18} and more sophisticated multitier $GW$+EDMFT~\cite{GW_EDMFT_PRM_Philipp20} have been reported to fail in predicting an insulating PM phase. 
Recently, SEET($GW$/ED) with an outer-loop self-consistency~\cite{Yeh20} was used to successfully open a gap for PM SrMnO$_{3}$ by adding on Mn $3d$ orbitals local self-energy corrections beyond the ones obtained in the sc$GW$. 
This is why we choose SrMnO$_{3}$ as an ideal strongly correlated PM insulator test case on which we will investigate GFCCSD.

\subsubsection{Comparison of impurity self-energies using ED and CCSD}
In Fig.~\ref{fig:SrMnO3_selfenergy_compare}, we plot the Mn $t_{2g}$, $e_{g}$, and O $p_{\pi}$, $p_{\sigma}$ impurity self-energy comparisons between the GFCCSD and ED impurity solvers, along with the double-counting correction coming from $GW$ restricted within the impurity subspace, in the first iteration of SEET. 
Same as for MnO, we use the same number of bath orbitals in GFCCSD and ED to eliminate any discrepancies due to different bath discretizations. 

For both Mn $t_{2g}$ and $e_{g}$ orbitals, GFCCSD reaches a qualitative agreement with ED for both the real and imaginary parts of the self-energy even though a tiny positive self-energy is observed for Mn $e_g$ orbitals. 
Quantitative deviations for the real part of self-energies appear in the high and low frequency region for Mn $t_{2g}$ and $e_{g}$, respectively. 
Note that the imaginary part of Mn $t_{2g}$ impurity self-energies from both ED and GFCCSD shows a divergent low-frequency behavior in contrast to the one of $GW$ double counting counterpart. 
This is likely an indication of metal-to-insulator transition where a similar divergence appears in the case of the 2D Hubbard model (as observed for example in Ref.~\cite{Kananenka15}).  Although a description of a realistic compound such as SrMnO$_3$ is much more complicated in the presence of the screening and hybridization from O $p$ orbitals and beyond, the above result implies that additional correlations added by SEET to Mn $t_{2g}$ orbitals  beyond the ones present in sc$GW$ are crucial for the opening of a gap in PM SrMnO$_{3}$. 

Next, we consider including non-perturbative self-energy corrections to the O $2p$ orbitals. 
For $p_{\pi}$ orbitals, despite the good agreement in the static part of the self-energy, large deviations are observed in the dynamic impurity self-energy between GFCCSD and ED solvers.
This observation is consistent with the one for the O $2p$ impurity in MnO. 
Moreover, for the imaginary part of the self-energy for the $p_{\pi}$ orbitals, we observe a positive and therefore a non-causal self-energy produced by the GFCCSD solver. This means that the correlations present in the $p_{\pi}$  orbitals are too strong and intractable at the level of singles and doubles excitations present in the GFCCSD approximation. 
In contrast, for O $p_{\sigma}$ orbitals, we observe an excellent agreement between the GFCCSD and ED solvers.

The comparisons of impurity self-energies between GFCCSD and ED solvers for various orbitals allow us to examine the strength of correlations among orbitals around $E_{F}$. We found that the O $p_{\sigma}$ orbitals are the least correlated, Mn $t_{2g}$, $e_{g}$ orbitals are modestly correlated, and O $p_{\pi}$ are the most strongly correlated ones.

\begin{table}[htp]
\begin{ruledtabular}
\begin{tabular}{c|c|p{6cm}}
Name & Imp & Description \\
\hline
A & 1 & Mn $t_{2g}$\\
B & 2 & Mn $t_{2g}$; Mn $e_g$ \\
C & 4 & Mn $t_{2g}$; Mn $e_g$; O $p_{\pi}$; O $p_{\sigma}$ \\
D & 3 & Mn $3d$; O $p_{\pi}$; O $p_{\sigma}$
\\
\end{tabular}
\end{ruledtabular}
\caption{
Different choices of impurities for the SrMnO$_{3}$ solid. Imp denotes the number of distinct, disjoint impurity problems. \label{tab:impurities_SrMnO3}}
\end{table}

\subsubsection{Local density of states}

In Fig.~\ref{fig:SrMnO3_PDOS}, we plot the local orbital-resolved DOS of SrMnO$_{3}$ in the PM phase calculated using SEET($GW$/CCSD) and SEET($GW$/ED) along with photoemission data~\cite{SrMnFeO3_expt_Kim10}. 
The impurity choices are listed in Table.~\ref{tab:impurities_SrMnO3}. 
According to Ref.~\cite{Yeh20}, the outer-loop self-consistency is crucial to open the gap for the PM phase of SrMnO$_{3}$ and, therefore, SEET with the outer-loop self-consistency is performed for both SEET($GW$/CCSD) and SEET($GW$/ED).

For impurity setup A, for details see Table~\ref{tab:impurities_SrMnO3}, despite problems of CCSD in finding a ground state of the impurity problem with proper number of particles (for details concerning energy ordering of different particle number present in the impurity see Sec.~\ref{sec:setupA_imp_search}) and small quantitative differences of impurity self-energies in the first iteration (see Fig.~\ref{fig:SrMnO3_selfenergy_compare} panel (a)), the self-consistent DOS from SEET($GW$/CCSD) is in a good agreement with DOS from SEET($GW$/ED).  

A non-perturbative treatment of only Mn $t_{2g}$ orbitals in SEET greatly suppress DOS at $E_{F}$ and opens a gap between Mn $t_{2g}$ + O $p_{\pi}$ valence bands and Mn $t_{2g}$ conduction bands, in contrast to sc$GW$ results~\cite{Yeh20} (which yield a metallic state). 
Such a result implies that treating local correlations of Mn $t_{2g}$ by a CCSD level solver is sufficient to open a gap. 

Although local correlation corrections beyond sc$GW$ that are added by SEET to Mn $t_{2g}$ orbitals open a gap, it has been shown that the gap is formed between Mn $t_{2g}$ + O $p_{\pi}$ valence states and Mn $e_{g}$ conduction band states~\cite{Bauernfeind18,Yeh20} and a non-perturbative treatment of Mn $e_{g}$ orbitals is necessary.
Consequently, we further treat Mn $e_{g}$ orbitals by SEET($GW$/CCSD) (setup B). 
SEET($GW$/CCSD) successfully pushes Mn $e_{g}$ conduction bands towards $E_{F}$ which is now ahead of the Mn $t_{2g}$ conduction band, similar as in SEET($GW$/ED).
However, in SEET($GW$/CCSD), the Mn $e_{g}$ conduction band is pushed too close to the Fermi level  such that the gap becomes too small when compared to results from SEET($GW$/ED) and photoemission data~\cite{SrMnFeO3_expt_Kim10}. 
This may due to the slight non-causality observed in Mn $e_{g}$ self-energy. 

Lastly, the non-perturbative treatment of O $2p$ orbitals is analyzed in setup C. While the SEET($GW$/CCSD) and SEET($GW$/ED) DOS looks similar, we have found that the particle sector searching for O $p_{\pi}$ is nontrivial in SEET($GW$/CCSD) and shows significant differences in comparison to the SEET($GW$/ED) case. For details see Sec.~\ref{sec:setupC_imp_search}.

\begin{figure}[htp]
\includegraphics[width=0.47\textwidth]{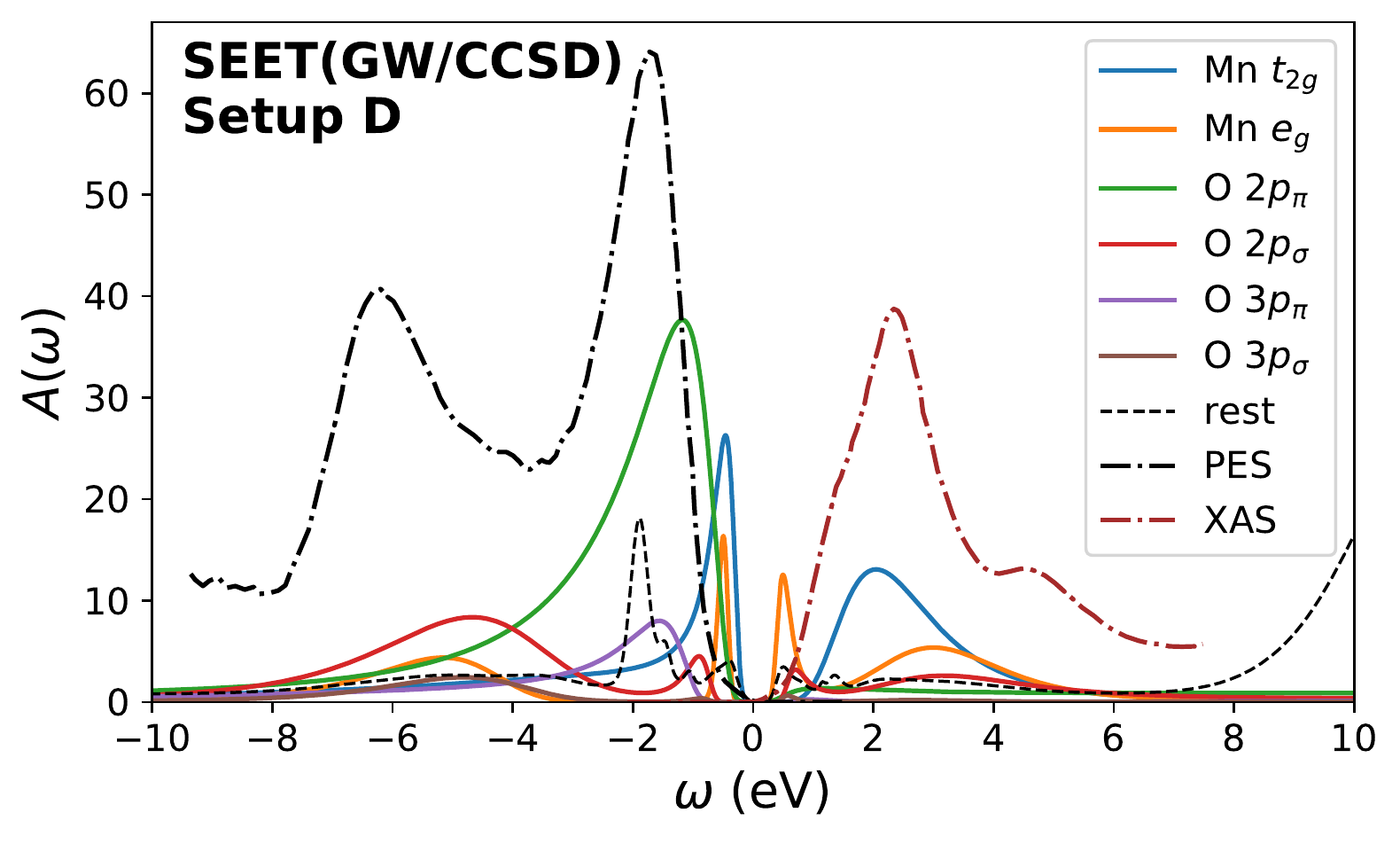}
\caption{Orbital-resolved local DOS of SrMnO$_{3}$ from SEET($GW$/CCSD) for impurity setup D (see Table.~\ref{tab:impurities_SrMnO3}). Dashed line are photoemission data from Ref.~\cite{SrMnFeO3_expt_Kim10}.}
\label{fig:SrMnO3_PDOS_D}
\end{figure}

To show the strength of GFCCSD impurity solver, in impurity setup D we further combine Mn $t_{2g}$ and $e_{g}$ orbitals into one bigger impurity in order to include inter-correlations between Mn $t_{2g}$ and $e_{g}$ orbitals. 
Note that setup D treats O $2p$ orbitals in the same manner as in impurity setup C where O $2p$ are split into $p_{\pi}$ and $p_{\sigma}$ impurities. 
Fig.~\ref{fig:SrMnO3_PDOS_D} shows the corresponding local DOS from SEET($GW$/CCSD). 
The most noticeable change when inter-correlations between $t_{2g}$ and $e_{g}$ are included is the shift of Mn $3d$ energy relative to O $p$ bands. 
In addition, Mn $e_{g}$ now has non-negligible contribution in the first valence peak from experiment. 
Unfortunately, we are not able to conclude whether this result is physical or simply an artifact from GFCCSD when enlarging the impurity size. 
Note that enlarging the impurity may lead to the rise of additional strong correlations resulting in an insufficiency of GFCCSD.
Therefore, a verification from calculations that include higher rank excitations is necessary and will be discussed in a future work. 

Although GFCCSD is able to treat impurities with more than couple tens of impurity orbitals, unfortunately for SrMnO$_3$, we were not able to converge SEET($GW$/CCSD) calculations with larger impurities e.g. containing both Mn $3d$ and O $2p$ orbitals combined together. This is due to the correlations which are intractable at the CCSD level. For such large impurities, we have observed a pronounced spin orbital symmetry breaking and instabilities in the particle sector search process. 

\begin{table*}[htp]
\begin{ruledtabular}
\begin{tabular}{c|c|c|c}
UHF & UCCSD & UCCSD(T) & ED \\
\hline
(9,6), (6,9) {\bf [-19.059]} & (9,6), (6,9) {\bf[-19.083]} & (8,7), (7,8) {\bf [-19.092]} & (9,6), (6,9), (8,7), (7,8) {\bf [-19.086]}\\
\hline
(9,7), (7,9) {\bf [-19.027]} & (8,7), (7,8) {\bf [-19.073]}& (9,6), (6,9) {\bf [-19.088]} &(8,7), (7,8) {\bf [-19.043]}\\
\hline
(8,7), (7,8) {\bf [-19.023]} & (8,6), (6,8) {\bf [-19.031]}& (8,6), (6,8) {\bf [-19.036]} &(8,7), (7,8) {\bf [-19.041]}\\
\end{tabular}
\end{ruledtabular}
\caption{Ordering of particle sectors for Mn $t_{2g}$ in SrMnO$_{3}$ found by UHF, UCCSD, UCCSD(T) and ED. The corresponding total energies (in a.u.) are denoted in square brackets. }
\label{tab:sectors_order_t2g}
\end{table*}

\begin{table*}[htp]
\begin{ruledtabular}
\begin{tabular}{c|c|c|c}
UHF & UCCSD & UCCSD(T) & ED \\
\hline
(8,5), (5,8), (6,8), (8,6) {\bf [-5.771]} & (6,6) {\bf [-5.855]} &(6,6) {\bf [-5.872]} & (6,6) {\bf [-5.866]} \\
\hline
(8,7), (7,8) {\bf [-5.765]} & (7,6), (6,7) {\bf [-5.846]} &(7,6), (6,7) {\bf [-5.861]} & (7,6), (6,7) {\bf [-5.855]} \\
\hline
(6,6) {\bf [-5.757]} & (8,6), (6,8) {\bf [-5.839]} &(7,7) {\bf [-5.849]} & (6,5), (5,6) {\bf [-5.848]} \\ 
\end{tabular}
\end{ruledtabular}
\caption{Ordering of particle sectors for O $p_{\pi}$ in SrMnO$_{3}$ found by UHF, UCCSD, UCCSD(T) and ED. The corresponding total energies (in a.u.) are denoted in square brackets.}
\label{tab:sectors_order_p_pi}
\end{table*}

\subsubsection{Particle number search for impurity in setup A}\label{sec:setupA_imp_search}
For impurity setup A, we first look at the energy ordering of particle sectors based on energies calculated using UHF, UCCSD, UCCSD(T), and ED in the first iteration of SEET (see Table~\ref{tab:sectors_order_t2g}). 

Note that in ED~\cite{ED_Sergei18} we observe a different degeneracy than in from UHF, UCCSD, and UCCSD(T). 
In ED, the ground state is a quartet state with the following z-component of spin $m_s=\{3/2,1/2,-1/2,-3/2\}$. 
The first excited state lies in the $(n_\alpha=8,n_\beta=7)$ sector is a doublet state with 32 mHartree higher in total energy than the ground state. 
On the other hand, UHF and UCCSD, respectively, yield lowest energy for $(n_\alpha=9,n_\beta=6)$ and $(n_\alpha=6,n_\beta=9)$ particle sectors equal to $m_s=\{3/2,-3/2\}$ values, thus breaking the degeneracy present in the quartet state.  
As more correlations are added UCCSD(T) finds $(n_\alpha=8,n_\beta=7)$ and $(n_\alpha=7,n_\beta=8)$ with $m_s=\{1/2,-1/2\}$ as the lowest state.
Note that the differences in the energies of different particle sectors are the result of the spin orbital symmetry breaking present in UCCSD and UCCSD(T). 
Fortunately for the impurity setup A, despite missing many components of $m_s$ in UCCSD and small quantitative differences of impurity self-energies in the first iteration (see Fig.~\ref{fig:SrMnO3_selfenergy_compare} panel (a)), self-consistent DOS from SEET($GW$/CCSD) is reasonably consistent with DOS from SEET($GW$/ED).

\subsubsection{Particle number search for impurity in setup C}\label{sec:setupC_imp_search}

The particle number search for impurity created using the O $2p_{\pi}$ orbitals is used in setup C of Fig.~\ref{fig:SrMnO3_PDOS}.
Table.~\ref{tab:sectors_order_p_pi} shows the orderings of different particle sectors based on total energy determined by UHF, UCCSD, UCCSD(T), and ED in the first iteration of SEET. 
As we mentioned before, particle sector search based on UHF frequently favors an unphysical ordering of states. In this case, the UHF ground state consists of open shell quartet with  $m_s=\{3/2, 1/2,-1/2,-3/2\}$ values  while the ED determined ground state is a singlet. 
UCCSD recovers the correct ground state and also the correct ordering of higher energy sectors as found in ED. 
Unfortunately, during self-consistent loop our particle-sector search based on UCCSD energy becomes unstable due to flipping of the energies of $(n_\alpha=6,n_\beta=6)$ and $((n_\alpha=6,n_\beta=7); (n_\alpha=7,n_\beta=6))$ particle sectors from iteration to iteration. 
Correlations beyond UCCSD are necessary to consistently distinguish the ground state $(n_\alpha=6,n_\beta=6)$  and the first excited states $((n_\alpha=6,n_\beta=7); (n_\alpha=7,n_\beta=6))$. 
Therefore, to stabilize the self-consistency procedure (similarly as in the case of MnO $2p$ impurity), we constrain the particle-sector space to states that contain the same number of $\alpha$ and $\beta$ electrons.

\section{Conclusions and discussion of the potential of  GFCC solvers}\label{sec:Conclusions}

We have investigated the performance of the GFCCSD solver for realistic impurity problems present in  AFM MnO and PM SrMnO$_{3}$ by analyzing impurity self-energies and local DOS for each of these compounds. Our \emph{ab initio} impurity Hamiltonians were constructed during the SEET($GW$/CCSD) self-consistency procedure. In this way, we examined impurity Hamiltonians present in realistic materials calculations avoiding possible simplifications that may be present in the low-energy models analyzed in the previous papers \cite{Avijit19, ZhuPRB19}.

Our work demonstrates that the GFCCSD solver is able to provide a satisfactory description for moderately correlated impurity problems. We observed that the self-energies from impurities containing $t_{2g}$ and $e_g$ orbitals of Mn were in an excellent agreement with the ones evaluated by ED. However, for impurities containing $2p$ orbitals of O both in the case of MnO and SrMnO$_3$, we observed significant discrepancies between ED and GFCCSD solvers. Consequently, we conclude that when correlations become stronger, higher order approximations in the CC theory are necessary for an excellent agreement with ED. This result is expected and completely supported by the experience gained in the quantum chemistry community with treating molecular systems at the CCSD level.

By limiting GFCC solver to the singles and doubles approximation, impurities with around hundred orbitals are very easily possible due to its polynomial scaling both for the parent UCCSD calculation and later at the Green's function construction stage. Note that this allows one to dramatically  extend both the number of impurity orbitals as well as the number of bath orbitals.
However, we also demonstrated that the singles and doubles approximation could lead to instabilities in the particle number search procedure when electron correlations are strong.
Consequently, for impurities containing a large number of $d$ and $p$ orbitals where multiple possibilities of degenerate orbitals leading to strong correlation effects exist, we advise caution when employing the GFCCSD solver.

In our opinion, the relative difficulty in searching for the particle number in the impurity problem is the most significant drawback of the GFCCSD solver since for zero temperature problems it may lead to the construction of Green's function corresponding to a ground state with a wrong particle number. 
Consequently, when performing SEET($GW$/CCSD) calculations with GFCCSD in zero-temperature limit, we are very careful to find an impurity ground state with a correct particle number.
For other approximate wavefunction-based solvers such as the ones based on RASCI~\cite{Zgid12}, it is possible to perform a finite temperature calculation where a Green's function is constructed using ground and excited states that are close in energy and can come from sectors with different number of particles. In this way multiple states are used and weighted with Boltzmann factors to construct a Green's function. In this procedure, even if the energy ordering of states is not perfect, the information from many states is retained mitigating small errors in the ordering of states. The same procedure cannot be applied to the CC method since for CC during the embedding self-consistency it is not possible in a completely automatic manner to generate couple of converged excited states for a given number of particles.


For larger impurity problems, especially where strong correlations are present, in order to meticulously analyze results for any realistic problem, especially in the absence of experimental data, one would like to know that the results from the GFCCSD solver are converged with respect to the excitation level. This is especially important when the impurity is enlarged since it is possible that the number of strongly correlated orbitals increases in the enlargement process making CCSD insufficient to illustrate all correlations and leading to wrong ordering of particle sectors. 
For additional discussion of the process of enlarging the impurity size evaluated in the GFCCSD solver, we refer the reader to Ref.~\cite{AS_TMO}.
This is why, in our opinion, to safely conduct many materials science calculations, we recommend checking the convergence of the CCSD ground state calculations  (which is the first step of the GFCCSD solver) by comparing CC energies or impurity orbital occupation numbers against CC results with higher excitations. If the results seem to be converged, then the result obtained at the GFCCSD level can be trusted and the ordering of particle sectors found is most likely sound.  Only such results should be used for physical interpretation of the results.
Ultimately, one would envision that a whole family of GFCC type of solvers can be developed with arbitrary high level of excitations leading to flexible tools where convergence with excitations can be always ensured.


In summary, we would like stress that at present the GFCCSD solver already yields impressive results for weakly and moderately correlated impurity problems. One of its biggest advantages is its ability to treat impurity problems with large number of orbitals and the possibility of enlarging both the correlated orbital space as well as the number of bath orbitals. We believe that including higher order excitations into the GFCC solver may bring a substantial improvement and enable fully predictable and systematically improvable {\em ab initio} calculations for novel compounds even in the absence of experimental data.

\acknowledgements
Ch-N. Y., A.S. and D. Z. acknowledge support of the Center for Scalable,
Predictive methods for Excitation and Correlated phenomena (SPEC), which is funded by the U.S. Department of Energy (DOE), Office of Science, Office of Basic
Energy Sciences, the Division of Chemical Sciences, Geosciences, and Biosciences.
S. I. is supported by the Simons foundation via the Simons Collaboration on the Many-Electron Problem This research partially used resources of the National Energy Research Scientific Computing Center (NERSC), a U. S. Department of Energy Office of Science User Facility operated under Contract No. DE-AC02-05CH11231.

\bibliographystyle{apsrev4-2}
\bibliography{refs}
\end{document}